%% file: LectureNotes.tex
\documentclass[envcountchap,sectrefs]{svmono}

\usepackage{mathptmx}
\usepackage{helvet}
\usepackage{courier}
\usepackage{type1cm}

\usepackage{makeidx}         
\usepackage{graphicx}        
\usepackage{multicol}        
\usepackage[bottom]{footmisc}

\usepackage{graphicx}
\usepackage{epsfig}

\usepackage{hyperref} 

\usepackage{bm}

\usepackage{amssymb}
\usepackage{amsfonts}
\usepackage{amsmath}

\newtheorem{Teorema}{Theorem}
\newtheorem{Lema}{Lemma}
\newtheorem{Propo}{Proposition}

\newtheorem{Definicao}{Definition}

\newtheorem{ex}{Example}
\newtheorem{ass}{Assumption}

\def\mathbi#1{\textbf{\em #1}}

\makeindex             

\begin{document}


\frontmatter

\include{first}

\include{preface}

\tableofcontents

\mainmatter
\include{chapterIntro_rev}

\include{chapterGraph_rev}

\include{chapterDynamics}

\include{chapterLinear1}

\include{chapterSync1}

\include{chapterClusterHyper}

\include{Conclusion}


\include{appendix_EDO}

\include{refe}
\printindex


\end{document}

%% file: first.tex
\begin{titlepage}

\vspace{2cm}
\begin{center}
{\Huge {Stability of Synchronized Motion in Complex Networks}}\\
\vspace{8.1 cm}
{\large \bf  Tiago Pereira} \\
Institute of Mathematical and Computer Sciences \\
University of São Paulo \\ 
tiago@icmc.usp.br
\vspace{1.3 cm}


\end{center}
\end{titlepage}

\thispagestyle{empty} 

\cleardoublepage




%% file: preface.tex
%
%

\preface


~ ~ These lectures are based on material which was presented in the Summer 
school at University of S\~ao Paulo, and in the winter school at the Federal University of ABC. The aim of this series is to introduce graduate students with a little background in the field to dynamical systems and network theory. 

Our goal is to give a succinct and self-contained description of the synchronized motion on 
networks of mutually coupled oscillators. We assume that the reader has basic knowledge of linear algebra and the theory of differential equations.

Usually, the stability criterion for the stability of synchronized motion is obtained in terms of 
Lyapunov exponents. We avoid treating the general case, for it would only bring further 
technicalities. We consider the fully diffusive case, which is amenable to treatment in terms 
of uniform contractions. This approach provides an interesting application of the stability theory 
and exposes the reader to a variety of concepts of applied mathematics, in particular, the theory 
of matrices and differential equations. More importantly, the approach provides a beautiful and rigorous, yet clear and concise, way to the important results. I expanded the initial notes slightly to include partial synchronization and synchronization in hypernetworks as an extra chapter, which was taught at the EU Marie Curie BeyondTheEdge project meeting.

The author has benefited from useful discussions with Murilo Baptista, Rafael Grissi, Kresimir Josic, Jeroen Lamb, Adilson  Motter, Ed Ott, Lou Pecora, Martin Rasmussen, Rafael Vilela, Eddie Nijholt, and Matthias Wolfrum. The author is indebted to Daniel Maia, Marcelo Reyes, and Alexei Veneziani for their critical reading of the manuscript. This work was partially supported by CNPq, FAPESP, the Leverhulme Trust grant RPG-279, and the EU Marie Curie IRSES Brazilian-European partnership in Dynamical Systems (BREUDS). We also acknowledge the support of the Humboldt Foundation via the Bessel Fellowship at the Weierstrass Institute for Analysis in Berlin.

\vspace{\baselineskip}
\begin{flushright}\noindent
S\~ao Carlos,\hfill {\it Tiago Pereira}\\
November  2024\hfill {\it }\\
\end{flushright}

%% file: chapterIntro_rev.tex
%
%
%
\motto{The art of doing mathematics consists in finding that special case which contains all the germs of generality. \\
\hspace{5.2cm}-- David Hilbert}

\chapter{Introduction}
\label{intro} 

~~ Real-world complex systems can be viewed and modeled as networks of interacting elements \cite{Newman,Albert1,Albert2}.  Examples range from geology \cite{Net3} and ecosystems \cite{Net4} to mathematical biology \cite{Net5} and neuroscience \cite{Bullmore}  as well as physics of neutrinos \cite{neutrinos} and superconductors \cite{StrogatzJunction}. Here we distinguish the structure of the network, the nature of the interaction, and the (isolated) dynamical behavior of individual elements. 

During the last fifty years, empirical studies of real complex systems have led to a deep understanding of the structure of networks, interaction properties, and isolated dynamics of individual elements, but a general comprehension of the resulting network dynamics remains largely elusive. 

Among the large variety of dynamical phenomena observed in complex networks, collective behavior is ubiquitous in real world networks and has proven to be essential to the functionality of such networks \cite{Net1,Net2,Fries,HubSync,RMP}. Synchronization is one of the most pervasive form collective behavior in complex systems of interacting components \cite{Strogatz,Kurths,Wu,Mech,Tiago,TiagoPhD_2006,DenizSync_Review}.
Along the riverbanks in some South Asian forests, whole swarms of fireflies will light up simultaneously in a spectacular synchronous 
flashing. Human hearts beat rhythmically because thousands of cells synchronize their activity \cite{Strogatz},  while thousands of 
neurons in the visual cortex synchronize their activity in response to specific stimuli \cite{WolfSinger}. Synchronization is rooted in human life, 
from the metabolic processes in our cells to the highest cognitive tasks \cite{Attention,Singer}.

Synchronization emerges from the collaboration and competition of many elements and has important consequences for all elements and network functioning. Synchronization is a multi-disciplinary discipline with a broad range of applications. Currently, the field experiences a vertiginous growth, and significant progress has already been made on various fronts.

Strikingly, in most realistic networked systems where synchronization is relevant, strong 
synchronization may also be related to pathological activities such as  seizures \cite{Ep,PRX} and Parkinson's disease \cite{parkinson} in neural networks, to extinction in ecology \cite{Extinction}, and social catastrophes in epidemic outbreaks \cite{Spread,Lai-Sang1,Lai-Sang2}. Of particular interest is how synchronization depends on various structural parameters such as degree distribution and spectral properties of the graph. 

In the mid-nineties Pecora and Carroll \cite{Pecora1} put forward a paradigmatic model of diffusively coupled identical oscillators on complex networks. They have shown that complex networks of identical nonlinear dynamical systems can globally synchronize despite exhibiting complicated dynamics at the level of individual elements. 

The analysis of synchronization in complex networks has benefited from advances in understanding the structure of complex networks \cite{Comb,Barrat,JEMS,StructuralGen,JNLS}. Barahona and Pecora \cite{Pecora2} have shown that well-connected 
networks -- with so-called small-world structure -- are easier to globally synchronize than regular networks. Motter and collaborators 
\cite{Motter} have shown that heterogeneity in the network structure hinders global synchronization. Later, it was shown that while heterogeneity may hinder global synchronization, it enhanced local cluster synchronization \cite{NonlinChimera,ZhengChaos,ZhengCMP}. Moreover, these findings can be extended to networks of non-identical oscillators \cite{PRL_T,Jaap}. These results form only the beginning of a proper understanding of the connections between network structure and the stability of global synchronization.

The approach put forward by Pecora and Carroll, which characterized the stability of  global synchronization, is based on elements of the theory of Lyapunov exponents \cite{LyapB}.  The characterization of stability via the theory of  Lyapunov exponents has many additional subtleties, in particular, when it comes to the persistence of stability under perturbations. A positive solution to the persistence problem requires the analysis of the so-called regularity condition, which is tricky and difficult to establish. 

We consider the fully diffusive case -- the coupling between oscillators depends only on their state difference. This model is amenable to full analytical treatment, and the stability analysis of the global synchronization is split into contributions coming solely  from the dynamics and from the network structure. The stability conditions in this case depend only on general properties of the oscillators and can be obtained analytically if one possesses knowledge of global properties of the dynamics, such as the boundedness of the trajectories. We establish the persistence under nonlinear perturbations and linear perturbations. Many conclusions  guide us  toward the ultimate goal of understanding more general collective behavior

%% file: chapterGraph_rev.tex
\motto{Can the existence of a mathematical entity be proved without definiing it?\\
\hspace{4.6cm} -- Jacques Hadamard}

\chapter{Graphs : Basic Definitions}

\section{Adjacency and Laplacian Matrices}

\hspace{.5cm} A network is a graph $G$ comprising a set of $n$ 
nodes (or vertices) connected by a set of $M$ links (or edges).  Graphs are the mathematical structures used to model pairwise relations between objects.  We shall often refer to the network topology, which is the layout pattern of  interconnections of the various elements. Topology can be considered as a virtual shape or structure of a network. 

The networks we consider here are  {\it simple} and {\it undirected}.  A network is called {\it simple} if the nodes do not have self-connections, and {\it undirected} if there is no distinction between the two vertices associated with each edge. A {\it path} in a graph is a sequence of connected (non-repeated) nodes.  From each node of a path, there is a link to the next node in the sequence. The length of a path is the number of links in the path. See further details in Ref. \cite{Graph}. 

\begin{figure}
\centerline{\hbox{\psfig{file=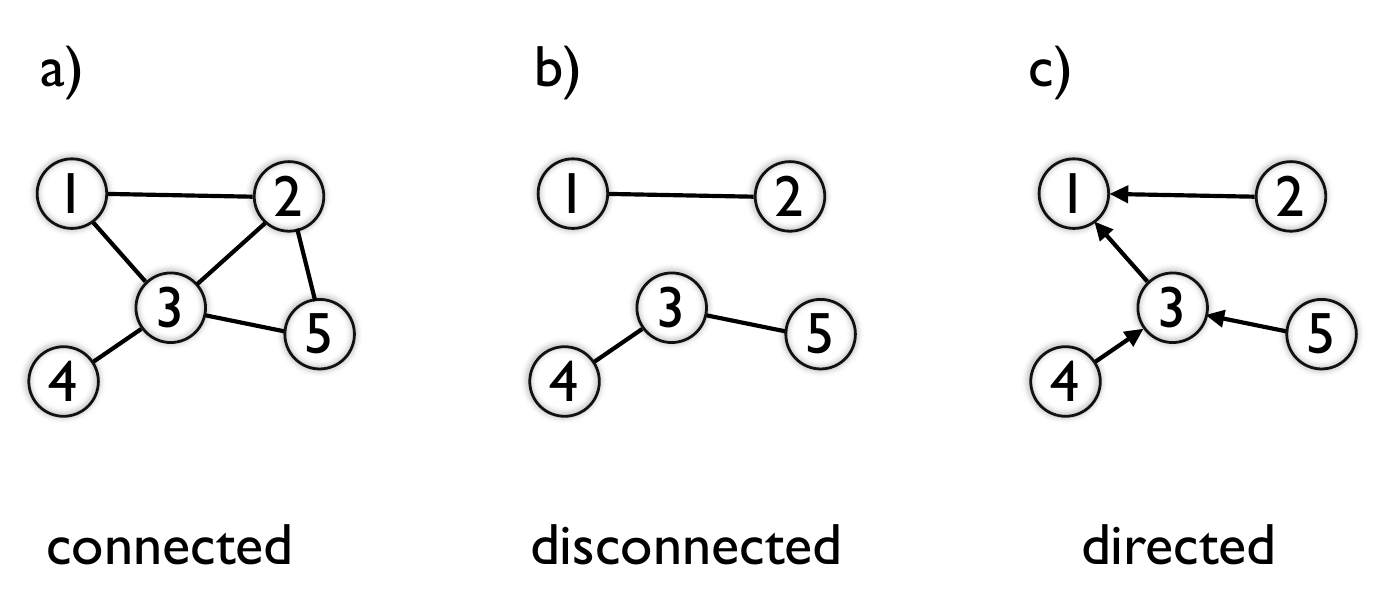,width=10.0cm}}}
\caption{Examples of undirected a) and b) and directed c) graphs. The diameter of graph a) is $d= 2$, hence, the graph is connected. The graph b) is disconnected, there is no path connecting the nodes 1 and 2 to the remaining nodes, the diameter is $d = \infty$. However, the graph has two connected components, the upper (1,2) with diameter $d=1$, and the lower nodes (3,4,5) with diameter $d = 2$. Graph c) is directed; the arrow tells the direction of the connection, so node 1 is reachable from node 2, but not the other way around. }
\label{GraphEx}
\end{figure}

For example, let's consider the network in Fig. \ref{GraphEx}a). Between the nodes $2$ and $4$ we have three paths $\{2,1,3,4\}$, $\{2,5,3,4\}$ and $\{2,3,4\}$. The first two have length $3$, and the last has length $2$. Therefore, the path $\{2,3,4\}$ is the shortest path between the node $2$ and $4$. 

The network {\it diameter} $d$ is the greatest length of the shortest path between any pair of vertices.  To find the diameter of a graph, first find the shortest path between each pair of vertices. The greatest length of any of these paths is the diameter of the graph. If we have an isolated node, that is, a node without any connections, then we say that the diameter is infinite. A network of finite diameter is called {\it connected}. 

A connected component of an undirected graph is a subgraph with finite diameter. The graph is called {\it directed} if it is not undirected. If the graph is directed, then there are two connected nodes, say, $u$ and $v$, such that $u$ is reachable from $v$, but $v$ is not reachable from $u$.  See Fig. \ref{GraphEx} for an illustration. 

The network may be described in terms of its {\it adjacency matrix}  $\mathbi{A}$, which encodes the topological information, and is defined as 
$$
A_{ij} = \left\{
\begin{array}{cc}
1 & \mbox{if } i \mbox{ and } j  \mbox{ are connected} \\
0 & \mbox{ otherwise }.
\end{array}
\right.
$$

An undirected graph has a symmetric adjacency matrix. The {\it degree} $k_i$ of the $i$th node  is the number of connections it receives, clearly 
$$
k_i = \sum_{j=1}^n A_{ij}.
$$ 

Another important matrix associated with the network is the combinatorial Laplacian matrix $\mathbi{L}$, defined as 
$$
L_{ij} = \left\{
\begin{array}{cc}
k_i & \mbox{ if } i=j \\
-1 & \mbox{if } i \mbox{ and } j  \mbox{ are connected} \\
0 & \mbox{ otherwise }.
\end{array}
\right.
$$

\begin{figure}[h]
\centerline{\hbox{\psfig{file=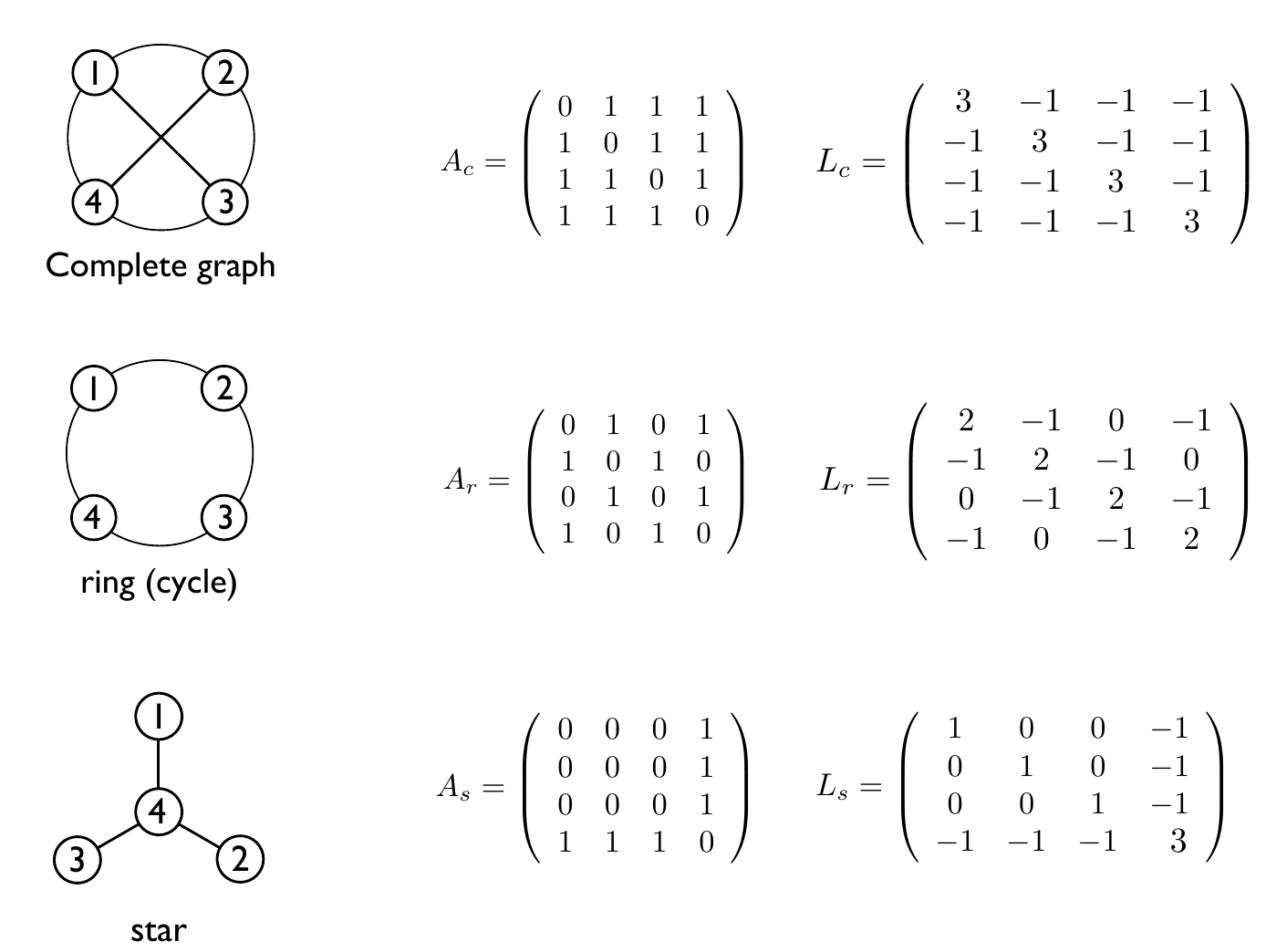,width=12.0cm}}}
\caption{Networks containing four nodes.  Their adjacency and Laplacian matrices are represented by $\mathbi{A}$ and $\mathbi{L}$. Further details can be found in Table \ref{table}. }
\label{Gexam}
\end{figure}

The Laplacian $\mathbi{L}$ is closely related to the adjacency matrix $\mathbi{A}$. In a compact form it reads 
$$
\mathbi{L} = \mathbi{D} -\mathbi{A}, 
$$
where $\mathbi{D} = $ diag$(k_1, \cdots, k_n)$ is the matrix of degrees. We depict in Fig. \ref{Gexam} distinct networks of size $4$ and their adjacency and Laplacian matrices. 
\noindent

\section{Spectral Properties of the Laplacian}

The eigenvalues and eigenvectors of $\mathbi{A}$ and $\mathbi{L}$ tell us a lot about the network structure. The eigenvalues of $\mathbi{L}$ for instance, are related to how well-connected the graph is and how fast a random walk on the graph could spread. In particular, the smallest nonzero eigenvalue of $\mathbi{L}$ will determine the synchronization properties of the network. Since the graph is undirected, the matrix $\mathbi{L}$ is symmetric, its eigenvalues are real,  and $\mathbi{L}$ has a complete set of orthonormal eigenvectors \ref{LA:ThmDiag}. The next result characterizes important properties of the Laplacian

\begin{Teorema}
Let $G$ be an undirected network and  $\mathbi{L}$ its associated Laplacian. Then: 
\begin{itemize}
\item[a)] \mathbi{L} has only real eigenvalues, 
\item[b)] $0$  is an eigenvalue and a corresponding 
eigenvector is $\mathbi{1} =  (1, 1, \cdots , 1)^*$, where $^*$ stands for the transpose. 
\item[c)] \mathbi{L} is positive semidefinite, its eigenvalues  enumerated 
in increasing order and repeated according to their multiplicity satisfy
$$
0= \lambda_1 \le \lambda_2 \le \cdots \le \lambda_n
$$ 
\item[d)] The multiplicity of $0$ as an eigenvalue of  $\mathbi{L}$ equals the 
number of connected components of G. 
\end{itemize}
\end{Teorema}

{\it Proof :}  The statement $a)$ follows from the fact that $\mathbi{L}$ is symmetric $\mathbi{L}=\mathbi{L}^*$, 
see Ap. \ref{LA} Theorem \ref{LA:ThmDiag}. To prove $b)$ consider the $\mathbi{1} =  (1, 1, \cdots , 1)^*$ and note that 
\begin{equation}
\label{Li}
(\mathbi{L 1})_i = \sum L_{ij} = k_i - \sum_j A_{ij} = 0
\end{equation}
Item $c)$ follows from the Gershgorin theorem, see Ap. \ref{LA} Theorem \ref{LA:ThmG}. The nontrivial conclusion $d)$ is 
one of the main properties of the spectrum. To prove the statement $d) $ we first note that if the graph 
$G$ has $r$ connected components $G_1, \cdots, G_r$, then is possible to represent 
$\mathbi{L}$ such that it splits into blocks $\mathbi{L}_1, \cdots , \mathbi{L}_r$. 

Let $m$ denote the multiplicity of $0$. Then Each $\mathbi{L}_i$ has an eigenvector $\mathbi{z}_i$ with $0$ as an eigenvalue. Note that $\mathbi{z}_i = (z_i^1, \cdots z_i^n)$ can be defined as $z_i^j$ is equal to 1 if $j$ belongs to the component $i$ and zero otherwise, hence $m \ge r$. It remains to show that any eigenvector $\mathbi{g}$ associated with $0$ is also constant. Assume that $\mathbi{g}$ is a non-constant eigenvector associated with $0$, and let $g_{\ell} >0$ be the largest entry of $\mathbi{g}$. Then 
\begin{eqnarray}
(\mathbi{L g})_{\ell} &=& \sum_j L_{\ell j} g_j  \nonumber \\
&=& \sum_j (k_{\ell} \delta_{\ell j} - A_{\ell j } ) g_j, \nonumber
\end{eqnarray}
since $\mathbi{g}$ is associated with the $0$ eigenvalue we have
$$
g_{\ell} = \frac{\sum_j A_{\ell j} g_j}{k_{\ell}}.
$$
This means that the value of the component $g_\ell$ is equal to the average of the values assigned to its neighbors. Hence $\mathbi{g}$ must be constant, which completes the proof. $\Box$

Therefore, $\lambda_2$ is bounded away from zero whenever the network is connected. The smallest non-zero eigenvalue is known as algebraic connectivity, and it is often called the Fiedler value.  The spectrum of the Laplacian is also related to some other topological invariants. One of the most interesting connections is its relation to the diameter, size, and degrees. 
\begin{Teorema} 
Let $G$ be a simple network of size $n$ and $L$ its associated Laplacian. Then: 
\begin{enumerate}
\item  \cite{Mohar} $ \lambda_2 \ge \displaystyle \frac{4}{nd} $
\item[]
\item \cite{Fiedler} $ \lambda_2 \le  \displaystyle \frac{n}{n-1} k_1 $
\end{enumerate}
\label{boundl}
\end{Teorema}

We will not present the proof of the Theorem here, however, it can be found in the references we provide in the theorem. We suggest the reader see further bounds on the spectrum of the Laplacian in Ref. \cite{Bojan}. Also, Ref. \cite{Bojan2} presents many applications of the Laplacian eigenvalues to diverse problems. One of the main goals in spectral graph theory is the obtain better bounds by having 
access to further information on the graphs.

For a fixed network size, the magnitude of $\lambda_2$ reflects how well-connected the graph is. Although the bounds given by Theorem \ref{boundl} are general, they can be tight for certain graphs.  For the ring, the lower bound on $\lambda_2$ is tight. This implies that  as the size increases -- consequently also its diameter -- $\lambda_2$  converges to zero, and the network becomes effectively disconnected. In sharp contrast, we find the star network. In this case, the upper bound in $i)$ is tight. The star diameter is equal to two, regardless of the size, and $\lambda_2=1$. See the table for the precise values. 
\begin{table}
\label{table}
\caption{Network of $n$ nodes. Examples of such networks are depicted in Fig. \ref{Gexam}}
\begin{center}
\begin{tabular}{ccccc}
\hline
Network 	&	$\lambda_2$   			&			$k_n $ & $k_1$ &	$D$  \\
\hline
\hline
Complete	&	 $ n $ &	$n-1$ & 	$n-1$	&	$1$ \\
~\\
ring	&	 $ \displaystyle 2  - 2\cos \left( \frac{2 \pi }{n}\right)$ &   $2$	& $2$	&		
$\begin{array}{c}
(n+1)/2 \mbox{ if  $n$ is odd} \\
n/2  \, \, \mbox{ \, \, \, \, if  $n$ is even} \\
\end{array}
$ 
\\
\\
Star & $\displaystyle 1$    &  $n-1$  & $1$  & 	2  \\
\\
\hline
\end{tabular}
\end{center}
\end{table}

The networks we encounter in real applications have a wilder connection structure. Typical examples are cortical networks, the Internet, power grids, and metabolic networks \cite{Newman}. These networks don't have a regular structure of connections, such as the ones presented in Fig. \ref{Gexam}. We say that the network is {\it complex} if it does not possess a regular connectivity structure. 

\begin{figure}
\centerline{\hbox{\psfig{file=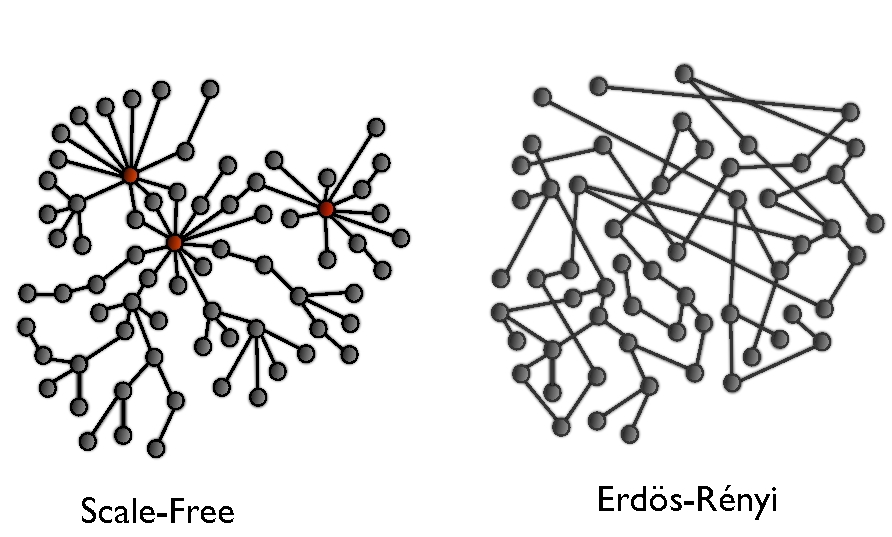,width=10.0cm}}}
\caption{Some examples of complex networks.}
\label{comnet}
\end{figure}

One of the goals is to understand the relation between the topological organization of the network and its functional relation, such as its collective motion. In Fig. \ref{comnet}, we depict two networks used to model real networks, namely the Barabasi-Albert and
the Erdos-Renyi Networks.  

The Erd\"os-R\'enyi network is generated by setting an edge between each pair of nodes with equal probability $p$, independently of the other edges. If $ p \gg \ln n / n$, then the network is almost surely connected, that is, as $N$ tends to infinity, the probability that a graph on $n$ vertices is connected tends to $1$. The degree is pretty homogeneous; almost surely every node has the same expected degree \cite{Comb}.

The Barabasi-Albert network possesses a great deal of heterogeneity in the nodes' degrees, while most nodes have only a few connections, some nodes, termed hubs, have many connections. These networks do not arise by chance alone.  The network is generated by means of the cumulative advantage principle -- the rich get richer. According to this process, a node with many links will have a higher probability of establishing new connections than a regular node.  The number of nodes of degree $k$ is proportional to $k^{-\beta}$. These networks are called  scale-free networks \cite{Newman}.  Many graphs arising in various real-world networks display a similar structure to the Barabasi-Albert network \cite{Albert1,Albert2}.

\noindent

%% file: chapterDynamics.tex
\motto{How can intuition deceive us at this point ? \\
\hspace{3cm} -- Henri Poincar\'e}

\chapter{Nonlinear Dynamics}

Let $D$ be an open simply connected subset of $\mathbb{R}^m$, $m \ge  1$, and let $\mathbi{f} \in C^r (D , \mathbb{R}^m)$ for some $r \ge 2$. We assume that the differential equation 
\begin{equation}
\frac{ d \mathbi{x}}{dt} = \mathbi{f}(\mathbi{x})
\label{nodeq}
\end{equation}
models the dynamics of a given system of interest. Now since $\mathbi{f}$ is differentiable the Picard-Lindel\"of  Theorem guarantees the existence of local solutions, see Ap. \ref{DE}  Theorem \ref{ThmPL}. We wish to guarantee that the solutions also exist globally. This requires further hypothesis on the behavior of the vector field. We are interested in systems that dissipate the volumes of $\mathbb{R}^m$ -- called dissipative systems.

\section{Dissipative Systems}

We say that  set  $\Omega \subset \mathbb{R}^m$ under the dynamics of Eq. (\ref{nodeq}) is positively invariant if the trajectories starting at the set never leave it in the future, that is, if $\mathbi{x}(t_0) \in \Omega$ then   $\mathbi{x}(t) \in \Omega$ for all $t \ge t_0$ Intuitively, it means that once the trajectory enters $\Omega$ it never leaves it again. The system is called dissipative if the solutions enter  a positively invariant set $\Omega \subset D$ in finite time. $\Omega$ is called the absorbing domain of the system. The existence of an absorbing domain guarantees that the solutions are bounded, hence, the extension results in  Ap. \ref{DE} 
Theorem \ref{ThmExt} assures the global existence of the solutions.

The question is then how to obtain the absorbing domains. Note that whenever $\mathbi{f}$ is nonlinear, finding the solutions of Eq. \ref{nodeq} can be a rather intricate problem. And usually we won't be able to do it analytically. So we need new machinery to address the problem on absorbing domains. A method by Lyapunov allows us to obtain such domains without finding the trajectories.  The technique infers the existence of the absorbing domains in relation to some properties
of a scalar function -- the Lyapunov function. 

We will study notions relative to connected nonempty subsets $\Omega$ of $\mathbb{R}^m$. A function $V: \mathbb{R}^m \rightarrow \mathbb{R}$ is said to be positive definite with respect to the set $B$ if $V(\mathbi{x})>0$ for all $\mathbi{x} \in \mathbb{R}^q\backslash \Omega$. 
It is radially unbounded  if 
$$
\lim_{\|\mathbi{x}\| \rightarrow \infty }V (\mathbi{x}) = \infty.
$$ 
Note that this condition guarantees that all level sets of $V$ are bounded. This fact plays a central role in the analysis.  We also define $V^{\prime} : \mathbb{R}^m \rightarrow \mathbb{R}$ as
$$
V^{ \prime}( \mathbi{x} ) = \nabla V (\mathbi{x})  \cdot \mathbi{f}(\mathbi{x}).
$$
where $\cdot$ denotes the Euclidean inner product. This definition agrees with the time derivative along the trajectories. That is, if $\mathbi{x}(t)$ is a solution of Eq. (\ref{nodeq}), then by the chain rule we have
$$
\frac{d V(\mathbi{x}(t))}{dt} = V^{\prime}(\mathbi{x}(t)).
$$
The main result is then the following

\begin{Teorema}[Lyapunov]
Let $V: \mathbb{R}^m \rightarrow \mathbb{R}$ be radially unbounded and positive definite with 
respect to the set $\Omega \subset D$. Assume that 
$$
V^{\prime}(\mathbi{x}) < 0 \mbox{  for all  }  \mathbi{x} \in \mathbb{R}^m \backslash \Omega
$$  
Then all trajectories of Eq. (\ref{nodeq}) eventually enter the set $\Omega$, 
in other words, the system is dissipative. 
\label{lyap}
\end{Teorema}

{\it Proof:} Note that for any trajectory $\mathbi{x}(t)$ in virtue of the fundamental theorem of the calculus
$$
V(\mathbi{x}(t)) - V(\mathbi{x}(s)) = \int_s^t V^{\prime}(\mathbi{x}(u)) du < 0. 
$$
So $V(\mathbi{x}(t)) < V(\mathbi{x}(s))$ for any $t > s$, and  $V$  
is decreasing along solutions and is radially unbounded, 
the level sets 
$$
S_a = \{ \mathbi{x} \in \mathbb{R}^m : V(\mathbi{x}) \le  a \}
$$ 
are positively invariant. Hence, the solutions are bounded,  and will lie in smaller level sets as time increases until the trajectory enters $\Omega$. It remains to show that once the solutions
lie in $\Omega$, they don't leave it. 

Suppose $\mathbi{x}(t)$ leaves $\Omega$ at $t_0$ and let $b = V( \mathbi{x}(t_0))$. The level set $S_b$  is closed, and there is  a ball $B_r(\mathbi{x}(t_0))$ such that $\mathbi{x}(t_0 +\varepsilon) \in B_r(\mathbi{x}(t_0)) \backslash S_b$ for some small $\varepsilon$.  Hence, $V(\mathbi{x}(t_0 + \varepsilon)) > V(\mathbi{x}(t_0))$ contradicting the fact that $V$ is decreasing along solutions. $\Box$

There are also converse Lyapunov theorems \cite{Liapunov}. Typically, if the system is dissipative (and has nice properties), then there exists a Lyapunov function. Although the above theorem is very useful, since we don't need knowledge of the trajectories, the drawback is the function $V$ itself. There is no recipe to obtain a function $V$ fulfilling all these properties. One could always try to guess the function, or go for a general form such as choosing a quadratic function $V$. We assume that the Lyapunov function is given. 

\begin{ass} \label{Dissip}
There exists a symmetric positive matrix \mbox{ $\mathbi{Q}$ }such that 
\end{ass}
$$
V(\mathbi{x}) = \frac{1}{2} (\mathbi{x} - \mathbi{a})^{*} \mathbi{Q} ( \mathbi{x} -\mathbi{a}).
$$
where $\mathbi{a} \in \mathbb{R}^m$. Consider the set $\Omega := \{ \mathbi{x} \in \mathbb{R}^m \, \, | \, \,  ( \mathbi{x} -\mathbi{a})^{\dagger} \mathbi{Q}  ( \mathbi{x} -\mathbi{a}) \le \rho^2 \}$, then 
$$
V^{\prime}(\mathbi{x}) < 0 \ , \forall \mathbi{x} \in \mathbb{R}^m \backslash \Omega.
$$

Under Assumption \ref{Dissip},  Theorem \ref{lyap} guarantees that $\Omega$ is positively invariant and that the trajectories of Eq. (\ref{nodeq}) eventually enter it. So, $\Omega$ is the absorbing domain of the problem. The solutions are, therefore, globally defined.

\section{Chaotic Systems}

Since the system Eq. (\ref{nodeq}) is dissipative, the solutions accumulate in a neighborhood of a bounded set $\Lambda \subset \Omega$.
The set $\Lambda$ is called an attractor. We focus on the situation where $\Lambda$ is a chaotic attractor. Now, the definition of a chaotic attractor is rather intricate -- there is even a general definition, the important properties for us are that solutions on the attractor are aperiodic, i.e., there is no $\tau \ge 0$ such that $\mathbi{x}(t) = \mathbi{x}(t + \tau)$, and the solutions exhibits sensitive dependence on initial conditions. Sensitive dependence on initial conditions means that nearby trajectories separate exponentially fast. 

If the system is chaotic, no matter how close two solutions start, they move apart when they are close to the attractor. Hence, arbitrarily small modifications of initial conditions typically lead to quite different states for large times.  This sensitive dependence on initial conditions is one of the main features of a chaotic system. Exponential divergence cannot go on forever, since the attractor is bounded, it is possible to show that the trajectories will come close together in the future \cite{nonlin}.

\subsection{Lorenz Model} \label{LorenzField}
The Lorenz model exhibits a chaotic dynamics \cite{Viana}. 
Using the notation 
$$
\mathbi{x} = 
\left(
\begin{array}{c}
x \\ y \\ z
\end{array}
\right),
$$
the Lorentz vector field reads
$$
\mathbi{f} (\mathbi{x}) = 
\left(
\begin{array}{c}
\sigma ( y -x )  \\ x ( r  - z) - y \\ -b z + xy 
\end{array}
\right)
$$
where we choose the classical parameter values $\sigma=10, r = 28, b = 8/3$. For these parameters, the Lorenz system fulfills our assumption \ref{Dissip} 
on dissipativity.

\begin{Propo} \label{OmeLor}
The trajectories of the Lorenz eventually enter  the absorbing domain 
\end{Propo}
$$
\Omega = \left\{ \mathbi{x} \in \mathbb{R}^3 \, \, : \, \, r x^2 + \sigma y^2 + \sigma (z - 2r )^2 < \frac{b^2  r^2}{b-1}  \right\}
$$

{\it Proof:} Consider the function 
$$
V(\mathbi{x}) =  (\mathbi{x} - \mathbi{a})^{\dagger} \mathbi{Q} (\mathbi{x} - \mathbi{a})
$$
where $\mathbi{a} = (0,0,2r)$ and $\mathbi{Q} = \mbox{diag}(r, \sigma, \sigma)$, note that the matrix is positive-definite. The goal is to find a bounded region $\Omega$ -- defined by means of a level set of $V$ -- such that $V^{\prime}<0$ in the exterior of $\Omega$ and then apply  Theorem \ref{lyap}. To this end, we compute the derivative, 
\begin{eqnarray}
V^{\prime}(\mathbi{x}) &=& 2 (\mathbi{x} - \mathbi{a})^{\dagger} \mathbi{Q} \mathbi{f}(\mathbi{x}) \nonumber \\
&=& 2 \sigma r x (y - x ) +\sigma y x (r - z) - \sigma y^2 - b\sigma z(z-2r)+\sigma(z-2r)xy \nonumber \\
&=& - 2 \sigma r x^2 - \sigma y^2 + \sigma y x (r - z) - b \sigma z (z-2r) - \sigma(r-z)xy \nonumber \\
&=& -2 \sigma \left[  rx^2 + y^2 + b (z - r )^2 - br^2 \right]. \nonumber
\end{eqnarray}

Consider the ellipsoid $E$ defined by $rx^2 + y^2 + b (z - r )^2 < br^2$, hence,  in the exterior of $E$ we have $V^{\prime}$. Now we take $c$ to be the largest value of $V$ in $E$, and we define $ \Omega = \{ \mathbi{x}\in \mathbb{R}^3 : V(\mathbi{x}) < c \} $.  The solutions will eventually enter $\Omega$ and remain inside since $V^{\prime} < 0$ in the exterior of $\Omega$,  and once the trajectory enters in $\Omega$ it never leaves the set. It remains to obtain the parameter $c$. This can be done by means of a Lagrange multiplier. After a computation -- see Appendix C of Ref. \cite{Sparow} -- we obtain $c = b^2 r^2/ (b-1)$ for $b\ge2$, and $\sigma \ge1$. $\Box$ 

\begin{figure}[h]
\centerline{\hbox{\psfig{file=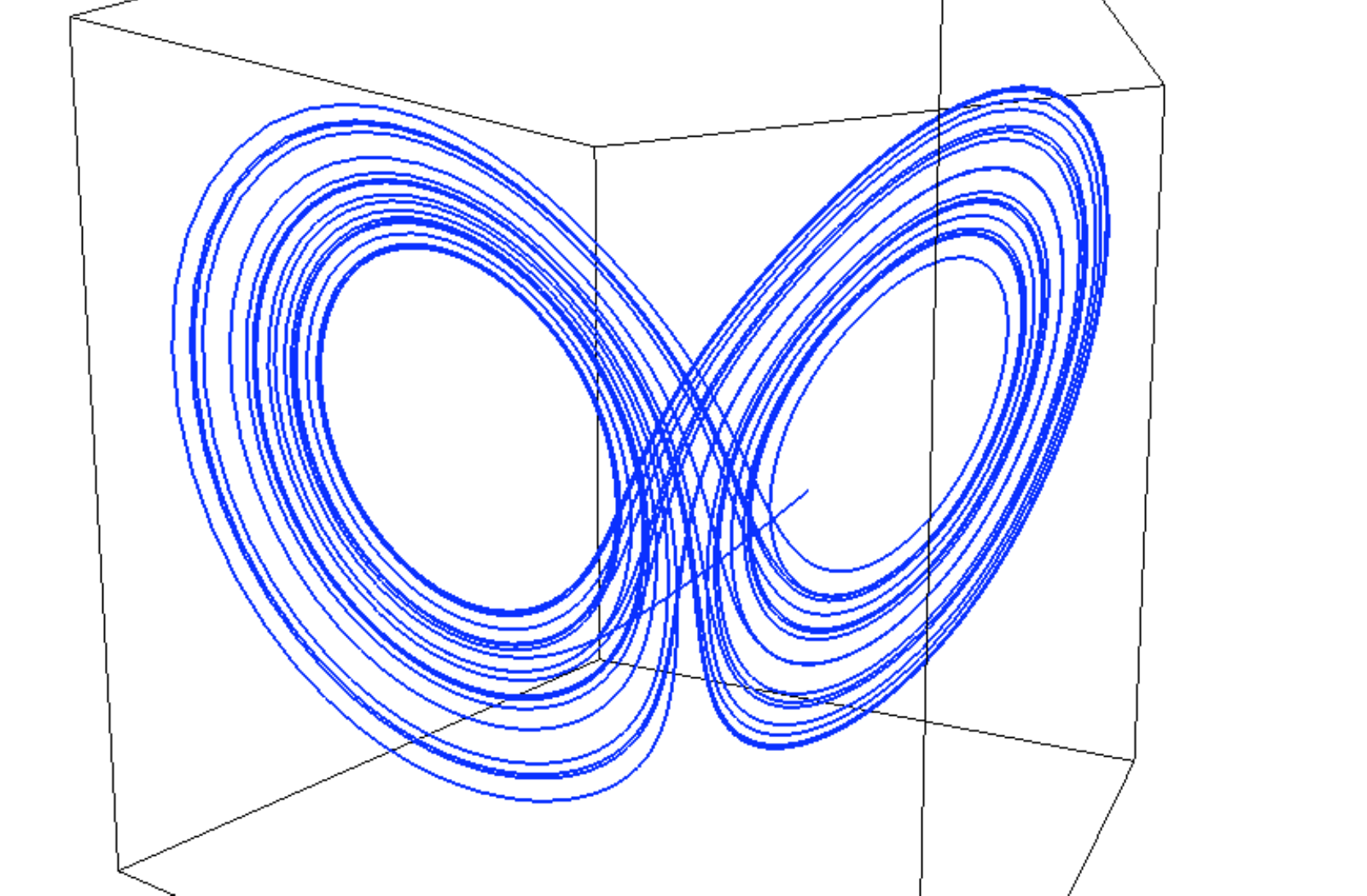,width=8.0cm}}}
\caption{The trajectories of the Lorenz system eventually enter an absorbing domain and accumulate on a chaotic attractor. This projection of an attractor resembles a butterfly -- the common name of the Lorenz attractor.}
\label{LoAt}
\end{figure}

Inside the absorbing set $\Omega$, the trajectory accumulates on the chaotic attractor. We have numerically integrated the Lorentz equations using a fourth-order Runge-Kutta, the initial conditions are $x(0)=-10$, $y(0)=10$, $z(0)=25$. We observe that the trajectory accumulates on the so-called Butterfly chaotic attractor \cite{Viana}, see Fig. \ref{LoAt}.

\begin{figure}[h]
\centerline{\hbox{\psfig{file=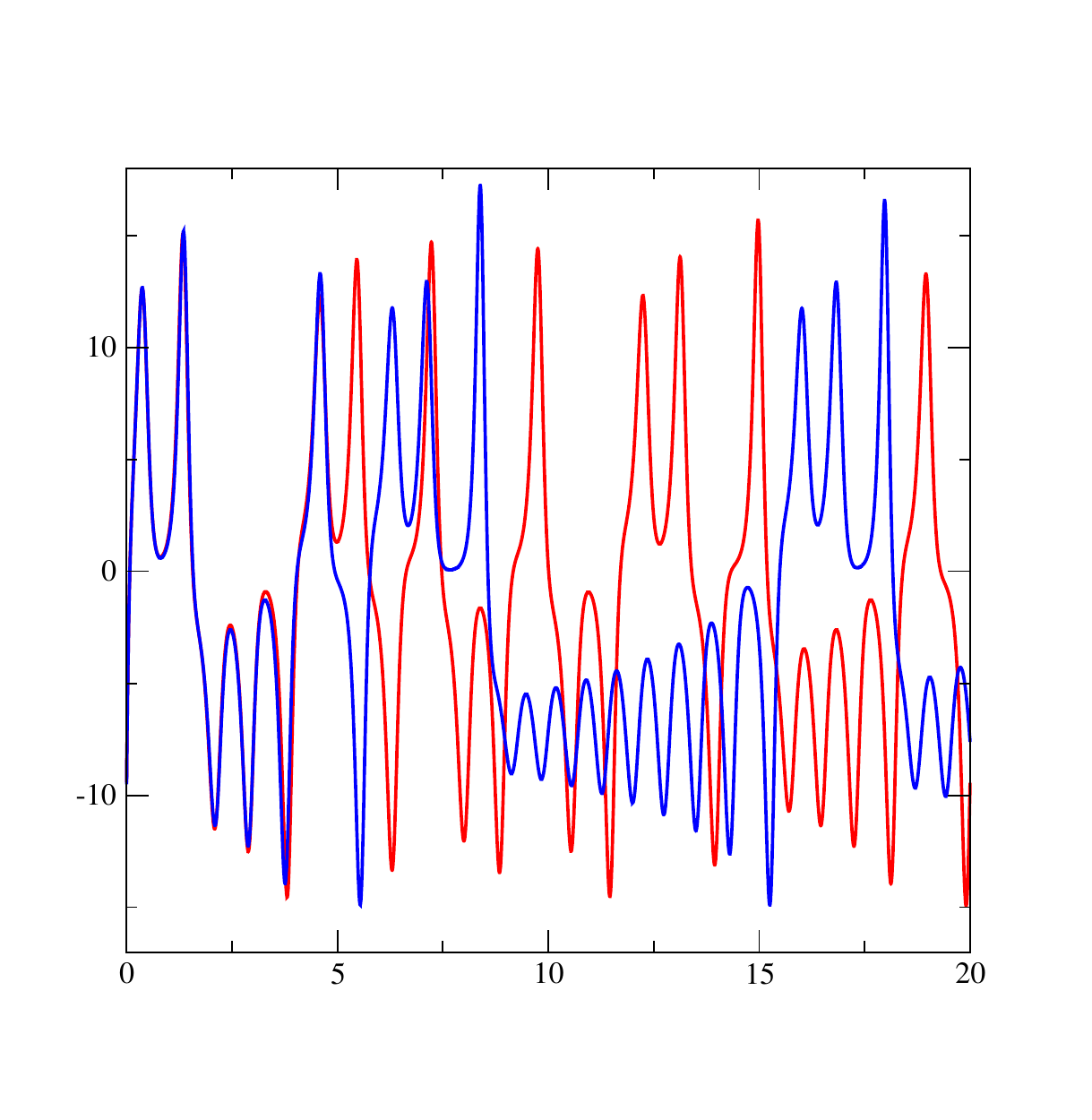,width=8.0cm}}}
\caption{Two distinct simulations of the time series $x(t)$ and $\tilde x(t)$ of the Lorentz systems.  
The difference between the trajectories is of $0.01$, however this small difference grows 
with time until a point where the difference is as large as the attractor itself.}
\label{Dif_Lo}
\end{figure}

Close to the attractor, nearby trajectories diverge. To see this phenomenon in a simulation, let us consider a distinct initial condition $\tilde{\mathbi{x}}(0) = 
(\tilde x(0) , \tilde y(0), \tilde z(0) )^{*}$. We consider $\tilde x(0)=-10.01$, $\tilde y(0)=10$, 
$\tilde z(0)=25$. Note that the initial difference $\| \mathbi{x}(0) - \tilde{\mathbi{x}}(0) \|_2 = 0.01$ 
becomes as large as the attractor size in a matter of 6 cycles, see  Fig. \ref{Dif_Lo}. 

\section{Diffusively Coupled Oscillators}\label{Nonlin:Diff}

We introduce now the network model. On top of each node of the network, we introduce a copy of the system Eq. (\ref{nodeq}). Then the influence that the neighbor $j$ exerts on the dynamics of  the node $i$ will be proportional to the difference of their state vector $\mathbi{x}_j(t) - \mathbi{x}_i(t)$. This type of coupling is called diffusive -- it tries to equate to state of the nodes. 

We label the nodes according to their degrees $k_n  \ge  \cdots \ge k_2  \ge k_1$, where $k_1$ and $k_n$ denote the minimal and maximal degree,  respectively. The dynamics of a network of $n$ identically diffusively coupled elements is described by 
\begin{eqnarray}
\frac{ d \mathbi{x}_i}{dt} &=& \mathbi{f}(\mathbi{x}_i) + \alpha \sum_{j=1}^n A_{ij}  
(\mathbi{x}_j - \mathbi{x}_i), \label{eqad} 
\end{eqnarray}
where  $\alpha$ is the overall coupling strength. In Eq. (\ref{eqad}) the coupling is given in terms of the adjacency matrix. 
We can also represent the coupled equations in terms of the network Laplacian.  Consider the coupling term
\begin{eqnarray}
\sum_{j=1}^n A_{ij}  (\mathbi{x}_j - \mathbi{x}_i) &=&  \sum_{j=1}^n A_{ij} \mathbi{x}_j - k_i \mathbi{x}_i \nonumber \\
&=&  \sum_{j=1}^n ( A_{ij}  - \delta_{ij} k_i )\mathbi{x}_j \nonumber 
\end{eqnarray}
where $\delta_{ij}$ is the Kronecker delta, and recalling that $L_{ij} = \delta_{ij} k_i - A_{ij}$ we obtain
Hence, the equations read
\begin{eqnarray}
\frac{ d \mathbi{x}_i}{dt} &=&\mathbi{f}(\mathbi{x}_i) - \alpha \sum_{j=1}^n L_{ij}  \mathbi{x}_j. \label{md1}
\end{eqnarray}

The dynamics of such a diffusive model can be intricate. Indeed, even if the isolated dynamics possesses 
a globally stable fixed point, the diffusive coupling can lead to the instability of the fixed points, and the 
systems can exhibit an oscillatory behavior. Please, see \cite{DiffusionDriven} for a discussion and further 
material. We will not focus on such a scenario of instability, but rather on how the diffusive coupling can lead 
to synchronization. 

Note that due to the diffusive nature of the coupling, if all oscillators start with the same initial condition, 
the coupling term vanishes identically. This ensures that the globally synchronized state 
$$
\mathbi{x}_1(t)=\mathbi{x}_2(t)= \cdots = \mathbi{x}_n(t) = \mathbi{s}(t), 
$$ 
is an invariant state for all coupling strengths $\alpha$.  The question is then the stability of synchronized 
solutions, which takes place due to coupling. Note that, if $\alpha=0$ the oscillators are decoupled, and 
Eq. (\ref{md1}) describes $n$ copies of the same oscillator with distinct initial conditions.  Since the 
chaotic behavior leads to a divergence of nearby trajectories, without coupling, any small perturbation 
on the globally synchronized motion will grow exponentially fast, and lead to distinct behavior 
between the node dynamics.

The correct way to see the invariant of globally synchronized motion is as follows. First consider 
$$
\mathbi{X} = \mbox{col}( \mathbi{x}_1 , \cdots , \mathbi{x}_n ),
$$
where col denotes the vectorization formed by stacking the column vectors $\mathbi{x}_i$ into a single 
column vector. Similarly 
$$
\mathbi{F}(\mathbi{X}) = \mbox{col}( \mathbi{f}(\mathbi{x}_1) , \cdots, \mathbi{f}( \mathbi{x}_n) ),
$$
then Eq. (\ref{md1}) can be arranged into a compact form
\begin{equation}
\label{comp}
\frac{ d \mathbi{X}}{dt} = \mathbi{F}(\mathbi{X}) - \alpha (\mathbi{L} \otimes \mathbi{I}_m ) \mathbi{X}
\end{equation}
where  $\otimes $ is the Kronecker product, see Appendix \ref{LA}. Let ${\Phi}( \cdot , t)$ be the flow of Eq. (\ref{comp}), the 
solution of the equation with initial condition $\mathbi{X}_0$ is given by $\mathbi{X}(t) = {\Phi}(\mathbi{X}_0, t)$.  
Consider the synchronization manifold
$$
\mathcal{M} = \{ \mathbi{x}_i \in \mathbb{R}^m : \mathbi{x}_i(t) = \mathbi{s}(t)  \mbox{ for  }  1\le i \le n\}, 
$$
then we have the following result 
\begin{Propo}
$\mathcal{M}$  is an invariant manifold under the flow ${\Phi}(\cdot , t)$
\end{Propo}

{\it Proof:} Recall that $\mathbi{1} \in \mathbb{R}^n$ is such that every component is equal to $1$.
Let $\mathbi{X}(t) = \mathbi{1} \otimes \mathbi{s}(t)$, 
note that 
$$
\frac{d \mathbi{X}(t)}{dt} = \mathbi{1} \otimes \frac{d \mathbi{s}(t)}{dt}.
$$
We claim that $\mathbi{X}(t)$ is a solution of the equations of motion. 
\begin{eqnarray}
\frac{d \mathbi{X}(t)}{dt} &=& \mathbi{F} ( \mathbi{X}(t) )  - \alpha (\mathbi{L} \otimes \mathbi{I}_m) \mathbi{X}(t) \nonumber \\
&=& \mathbi{F} ( \mathbi{1} \otimes \mathbi{s}(t) )  - \alpha ( \mathbi{L} \otimes \mathbi{I}_m ) \mathbi{1} \otimes \mathbi{s}(t) \nonumber \\
&=&\mathbi{1} \otimes \mathbi{f}( \mathbi{s}(t)) \nonumber 
\end{eqnarray}
where in the last passage we used Theorem (\label{Thm:KroPod})  together with \mathbi{L}  \mathbi{1} = 0 and $\mathbi{F}(\mathbi{1} \otimes \mathbi{s}(t)) = \mathbi{1} \otimes \mathbi{f}(\mathbi{s}(t))$. By the Picard-Lindel\"of Theorem \ref{ThmPL} we have that that  $\mathbi{X}(t) = {\Phi}(\mathbi{1} \otimes \mathbi{s}(0), t) \in \mathcal{M}$ for all $t$. $\Box$

If the oscillators have the same initial condition, their evolution will be exactly the same
forward in time, no matter the value of the coupling strength. 

In the above result, we have looked at the network not as a coupled equation but as a single system in the full 
state space $\mathbb{R}^{m n}$. We prefer to keep the picture of coupled oscillators. These pictures are 
equivalent, and we interchange them whenever it suits our purposes. The important questions are
\begin{description}
\item\mathbi{ Boundedness} of the solutions $\mathbi{x}_i(t)$. 
\item{\bf Stability} of the globally synchronized state (synchronization manifold). 
\end{description}

We wish to address the local stability of the globally synchronized state. That is, if all trajectories start close 
together $\| \mathbi{x}_i (0) - \mathbi{x}_j(0) \| \le \varepsilon$, for any $i$ and $j$ and some small $\varepsilon$, 
would they converge to $\mathcal{M}$, in other words, would 
$$
\lim_{t \rightarrow \infty}\|\mathbi{x}_i(t) - \mathbi{x}_j(t) \| = 0
$$
or would the trajectories split apart?  The goal of the remaining exposition is to provide positive answers 
to these questions. To this end, we review some fundamental results needed to address such points.

%% file: chapterLinear1.tex
\motto{The more you know, the less sure you are. \\
\hspace{3cm} -- Voltaire}

\chapter{Linear Differential Equations}

The question concerning the local stability of a given  trajectory $\mathbi{s}(t)$ leads to the stability analysis
of the trivial solution of a nonautonomous linear differential equation. The analysis of the dynamics in a neighborhood of the solutions is performed by using the variational equation. The trajectory \mathbi{s}(t) is 
stable when the stability of the trivial solution of the variational equation is preserved under small 
perturbations.

\section{First Variational Equation}

Let $\mathbi{y} (0)$ be close to $\mathbi{s}(0)$. Each of these 
distinct points has its behavior determined by the equation of motion Eq. (\ref{nodeq}). 
We can follow the dynamics of the difference 
$$
\mathbi{z} (t) =\mathbi{y}(t) - \mathbi{s}(t)
$$
which leads to the variational equations governing its evolution
\begin{eqnarray}
\frac{ d \mathbi{z} (t)}{dt}  &=&  \mathbi{f}(\mathbi{y}(t))  - \mathbi{f}(\mathbi{s}(t)) \nonumber \\
& = & \mathbi{f}(\mathbi{s}(t) + \mathbi{z} (t))  - \mathbi{f}(\mathbi{s}(t)), \nonumber 
\end{eqnarray}
now since $\| \mathbi{z} \|$ is sufficiently small we may expand the function $\mathbi{f}$ in 
Taylor series 
$$
\mathbi{f}(\mathbi{s}(t) + \mathbi{z} (t))  = \mathbi{f}(\mathbi{s}(t)) + D\mathbi{f}(\mathbi{s}(t)) \mathbi{z} (t) + \mathbi{R}(\mathbi{z}(t))
$$
where $D\mathbi{f}(\mathbi{s}(t))$ along the trajectory $\mathbi{s}(t)$, and by the Lagrange 
theorem \cite{CalculusRn}
$$
\| \mathbi{R}(\mathbi{z}(t)) \| = O(\| \mathbi{z}(t) \|^2).
$$ 
Truncating the evolution equation of $\mathbi{z}$, up  to the first order,  we obtain the 
first variational equation
\begin{eqnarray}
\frac{ d \mathbi{z}}{dt} = D\mathbi{f}(\mathbi{s}(t)) \mathbi{z}. \nonumber
\end{eqnarray}
Note that the above equation is non-autonomous and linear. Moreover, since $\mathbi{s}(t)$ lies in a 
compact set and $\mathbi{f}$ is continuously differentiable, by Weierstrass Theorem \cite{CalculusRn}, 
$D\mathbi{f}(\mathbi{s}(t))$ is a bounded matrix function. If $\| \mathbi{z} (t) \| \rightarrow 0$ the two 
distinct solutions converge to each other and have an identical evolution.  

The first variational equation plays a fundamental role to tackling the local stability problem. Suppose 
that somehow we have succeeded to demonstrate that the trivial solution of the first 
variational equation is stable. Note that this does not completely solve our problem, because the Taylor 
remainder acts as a perturbation of the trivial solution. Hence, to guarantee that the problem can be solved 
in terms of the variational equation we must also obtain conditions on the persistence of the stability of  
trivial solution under small perturbation. There is a beautiful and simple, yet general, criterion based on 
uniform contractions.  We follow closely the exposition in Ref. \cite{Coppel,LinearSys}.

\section{Stability of Trivial Solutions}

Consider the linear differential equation 
\begin{equation}
\frac{ d \mathbi{x}}{dt} = \mathbi{U}(t)\mathbi{x}
\label{eqlin}
\end{equation}
where $\mathbi{U}(t)$ is a continuous bounded linear operator on $\mathbb{R}^q$ for each $t \ge 0$.

The point $\mathbi{x} \equiv \mathbi{0}$ is an equilibrium point of the equation Eq. (\ref{eqlin}). 
Loosely speaking, we say an equilibrium point is locally stable if the initial conditions are in a 
neighborhood of zero solution remain close to it for all time. The zero solution is said to be locally 
asymptotically stable if it is locally stable and, furthermore, all solutions starting near 
$\mathbi{0}$ tend towards it as $t\rightarrow \infty$.
 
The time dependence in Eq. (\ref{eqlin}) introduces of additional subtleties \cite{rasmussen}. Therefore, we want 
to state some precise definitions of stability 

\begin{Definicao}[Stability in the sense of Lyapunov]
The equilibrium point $\mathbi{x}^*=0$  is stable in the sense of Lyapunov at $t=t_0$ if for any
$\varepsilon >0$ there exists a $\delta(t_0,\varepsilon)>0$ such that
$$
\| \mathbi{x}(t_0)\| < \delta	\Rightarrow  \| \mathbi{x}(t) \| < \varepsilon,	\, \ \forall t \ge  t_0
$$
\end{Definicao}
Lyapunov stability is a very mild requirement on equilibrium points. In particular, it does not require that 
trajectories starting close to the origin tend to the origin asymptotically. Also, stability is defined at a time 
instant $t_0$. Uniform stability is a concept which guarantees that the equilibrium point is not losing stability. 
We insist that for a uniformly stable equilibrium point $\mathbi{x}^*$, $\delta$ in the Definition 4.1 not be a 
function of $t_0$, so that equation may hold for all $t_0$. Asymptotic stability is made precise in the 
following definition:

\begin{Definicao}[Asymptotic stability]
An equilibrium point $\mathbi{x}^* = 0$ is asymptotically stable at $t = t_0$ if 
\begin{enumerate}
\item $\mathbi{x}^* = 0$ is stable, and 
\item $\mathbi{x}^* = 0$ is locally attractive; i.e., there exists $\delta(t_0)$ such that 
$$
\| \mathbi{x}(t_0) \| < \delta	 \ \Rightarrow \ \lim_{t \rightarrow \infty} \mathbi{x}(t) = \mathbi{0}
$$
\end{enumerate}
\end{Definicao}

\begin{Definicao}[Uniform asymptotic stability]
An equilibrium point $\mathbi{x}^* = 0$ is uniform asymptotic stability if 
\begin{enumerate}
\item $\mathbi{x}^* = 0$ is asymptotically stable, and
\item there exists $\delta_0$ independent of $t_0$ 
for which equation holds. Further, it is required that the convergence is uniform. That is, 
for each $\varepsilon>0$ a corresponding $T = T(\varepsilon) > 0$ such that if 
$\| \mathbi{x}(s) \| \le \delta_0$ for some $s \ge 0$ then $\| \mathbi{x}(t) \| < \varepsilon$ for all
$t \ge s + T$. 
\end{enumerate}
\end{Definicao}

We shall focus on the concept of uniform asymptotic stability. To this end, we wish to 
express the solutions of the linear equation in a closed form. The theory of differential 
equations guarantees that the unique solution of the above equation 
can be written in the form
$$
\mathbi{x}(t) = \mathbi{T}(t,s) \mathbi{x}(s)
$$
where $\mathbi{T}(t,s)$ is the associated evolution operator \cite{LinearSys}. The evolution operator 
satisfies the following properties
\begin{eqnarray}
\mathbi{T}(t,s) \mathbi{T}(s,u) &=& \mathbi{T}(t,u) \nonumber \\
\mathbi{T}(t,s) \mathbi{T}(s,t) &=& \mathbi{I}_m. \nonumber
\end{eqnarray}

The following concept plays a major role in these lectures

\begin{Definicao}
Let $\mathbi{T}(t,s)$ be the evolution operator associated with Eq. (\ref{eqlin}).
$\mathbi{T}(t,s)$ is said to be a uniform contraction if 
$$
\|\mathbi{T}(t,s) \| \le K e^{-\eta (t-s)}.
$$
where $K$ and $\eta$ are positive constants. 
\end{Definicao}

Some examples of evolution operators and uniform contractions are 

\begin{ex}\label{Ex:Linear}
If $\mathbi{U}$ is a constant matrix, then Eq. (\ref{eqlin}) is autonomous, and the 
fundamental matrix reads
$$
\mathbi{T}(t,s) = e^{(t-s) \mathbi{U}},
$$
$\mathbi{T}(t,s)$ has  a uniform contraction if, and only if all its eigenvalues have negative real part. 
\end{ex}

\begin{ex}
Consider the scalar differential equation
$$
x^{\prime} = \{ \sin \log (t+1) + \cos \log (t+1) - b \} x,
$$
the evolution operator reads 
$$
T(t,s) = exp\{ - b (t-s) +  (t+1) \sin \log (t+1) - (s+1) \sin \log (s+1) \}
$$
Then following holds for the equilibrium point $x=0$
\begin{enumerate}
\item[i)] If $b < 1$, the equilibrium  is unstable.
\item[ii)] If $b = 1$, the equilibrium is stable but not uniformly stable.
\item[iii)] If $ 1 < b < \sqrt{2}$, the equilibrium is asymptotically stable but not uniformly stable 
or uniformly asymptotically stable.
\item[iv)] If $b = \sqrt{2}$, the equilibrium is asymptotically stable. Though it is uniformly stable, it is not 
uniformly asymptotically stable.
\item[v)] If $b > \sqrt{2}$, the equilibrium is uniformly asymptotically stable.
\end{enumerate}
\end{ex}

We will show that the trivial solution of Eq. \ref {eqlin} is uniformly asymptotically stable if, and only if,
the evolution operator is a uniform contraction, that is, the solutions converge converges 
exponentially fast to zero.

\begin{Teorema} \label{UniCon}
The trivial solution of Eq. (\ref{eqlin}) is uniformly asymptotic stable if, and only if the evolution operator
is a uniform contraction. 
\end{Teorema}

{\it Proof:} First suppose the evolution operator is a uniform contraction then 
\begin{eqnarray}
\| \mathbi{x}(t) \| & = &  \| \mathbi{T}(t,s) \mathbi{x}(s) \| \nonumber \\
& \le &  \| \mathbi{T}(t,s) \| \|  \mathbi{x}(s) \| \nonumber \\
& \le &  K e^{ - \alpha (t-s) } \|  \mathbi{x}(s) \|. \nonumber 
\end{eqnarray}
Now let $\varepsilon > 0$ be given, clearly if $t > T$, where $T = T(\varepsilon)$ is 
large enough then the $\| \mathbi{x}(t) \| \le \varepsilon.$ Let $\| \mathbi{x}(s)\| \le \delta$, we 
obtain $ \| \mathbi{x}(t) \|  \le  K e^{ - \alpha (t-s) } \delta < \varepsilon, $ 
which implies that 
$$
T=T(\varepsilon) = \frac{1}{\alpha} \ln \frac{\delta K}{\varepsilon},
$$
completing the first part. 

To prove the converse, we assume that the trivial solution is uniformly asymptotically stable. Then 
there is $\delta$ such that for any  $\varepsilon$ and $T = T(\varepsilon)$ such that 
for any $\| \mathbi{x}(s) \| \le \delta $ we have
$$
\| \mathbi{x}(t) \| \le \varepsilon,
$$  
for any $t \ge s + T$. Now take $\varepsilon = \delta / k$, and consider the sequence
$t_n = s + nT$. 

Note that 
$$
\| \mathbi{T}(t,s) \mathbi{x}(s) \|  \le \frac{\delta}{k},
$$  
for any $\| \mathbi{x}(s) \| / \delta \le 1 $, we have the following bound for the norm
$$
\| \mathbi{T}(t,s) \| = \sup_{ \| \mathbi{u} \| \le 1 } \| \mathbi{T}(t,s) \mathbi{u} \|  \le \frac{1}{k}. 
$$  
Remember that $  \mathbi{T}(t,u)  \mathbi{T}(u,s) =  \mathbi{T}(t,s)$. Hence, 
\begin{eqnarray}
\| \mathbi{T}(t_2,s) \| &=& \|   \mathbi{T}(s + 2T,s + T)   \mathbi{T}(s + T,s)  \| \nonumber \\
&\le & \|   \mathbi{T}(s + 2T,s + T) \| \|  \mathbi{T}(s + T,s)  \| \nonumber \\
& \le & \frac{1}{k^2}.  \nonumber
\end{eqnarray}

Likewise, by induction
$$
\| \mathbi{T}(t_n,s) \| \le \frac{1}{k^n}, 
$$
take $ \alpha  = \ln k / T$, therefore, 
$$
\| \mathbi{T}(t_n,s) \| \le  e^{-\alpha (t_n-s)}. 
$$

Consider the general case $t = s + u + nT$, where $0\le u < T$, then the same bound 
holds 
\begin{eqnarray}
\| \mathbi{T}(t,s) \| & \le &  e^{ - nT \alpha}   \nonumber \\
& \le &  K e^{ - (t - s) \alpha},   \nonumber
\end{eqnarray}
where $K \le  e^{\alpha T}$, and we conclude the desired result. $\Box$

\section{Uniform Contractions and Their Persistence}

The uniform contractions have a rather important roughness property, they 
are not destroyed under perturbations of the linear equations. 

\begin{Propo}\label{Rough}
Suppose $\mathbi{U}(t)$ is a continuous matrix function on $\mathbb{R}_+$ and consider 
Eq. (\ref{eqlin}). Assume the fundamental matrix $
\mathbi{T}(t,s)$ has a uniform contraction. Consider a continuous matrix function $\mathbi{V}(t)$ 
satisfying 
$$
\sup_{t \ge 0} \| \mathbi{V} (t) \| =  \delta \le \frac{\eta}{K} 
$$
then the evolution operator $\hat T(t,s) $ of the perturbed equation 
$$
\frac{ d\mathbi{y}}{dt} = [ \mathbi{U}(t) + \mathbi{V}(t) ]\mathbi{y},
$$
also has a uniform contraction satisfying
$$
\| \hat{\mathbi{T}}(t,s) \| \le K e^{-\gamma (t-s)},
$$
where $\gamma = \eta - \delta K$.
\end{Propo}

{\it Proof:} Let us start by noting that the evolution operator $\mathbi{T}(t,s)$ also 
satisfies  the differential equation of the unperturbed problem
$$
\frac{d}{dt}\mathbi{T}(t,s) = \mathbi{U}(t) \mathbi{T}(t,s), 
$$
The evolution operator $\hat{\mathbi{T}}$ can be obtain by the variation of parameter, 
see Ap. \ref{DE} Theorem \ref{ThmVP}. So,  
$$
\hat{\mathbi{T}}(t,s) = \mathbi{T}(t,s) + \int_s^t \mathbi{T}(t,u) \mathbi{ V}(u) \hat{\mathbi{T}}(u,s)du,
$$
using the induce norm, for $t \ge s$, 
$$
\| \hat{\mathbi{T}}(t,s) \| \le K e^{- \eta (t-s)} + \delta K \int_s^t e^{-\eta (t-u)}  \| \hat{\mathbi{T}}(u,s) \|du.
$$
Let us introduce the scalar function $w(u) = e^{-\eta (t-u)} | \hat{\mathbi{T}}(t,s) |$, then 
$$
w(t) \le K w(s) + K \delta \int_s^t  w(u)du, 
$$
for all $t\ge s$. Now we can use the Gronwall's inequality to estimate $w(t)$, see Ap. \ref{DE} Theorem \ref{ThmGI}, this implies
$$
w(t) \le  K w(s) e^{ \delta K ( t-s )},
$$
consequently
$$
\| \hat{\mathbi{T}}(t,s) \| \le K e^{\left( \eta - K \delta \right) (t-s) }.
$$
$\Box$

The roughness property of uniform contraction does the job and guarantees
that the stability of the trivial solution is maintained. The question now turns to 
how to obtain a criterion for uniform contractions. There are various criteria, and 
the following suits our purposes

\begin{Lema}[Principle of Linearization]
Assume the the fundamental matrix $\mathbi{T}(t,s)$ of Eq. (\ref{eqlin}) has a uniform contraction.
Consider the perturbed equation 
$$
\frac{ d \mathbi{y}}{dt} = \mathbi{U} (t) \mathbi{y} + \mathbi{R}(\mathbi{y}),
$$
and assume that 
$$
\| \mathbi{R} (\mathbi{y}) \| \le M \| \mathbi{y} \|^{1+c},
$$
for some $c> 0$. Then the origin is exponentially asymptotically stable. 
\label{PL}
\end{Lema}

{\it Proof:} Note that we can write $\| \mathbi{R} (\mathbi{y}) \| \le K \| \mathbi{y} \|^{1}$, 
where $K = M \| \mathbi{y} \|$. Now given a neighborhood of the trivial solution $\| \mathbi{y}\|\le \delta$
is possible to control $K \le \varepsilon$. Applying the previous Proposition \ref{Rough} we conclude
the result.  $\Box$

This result can be used to prove that if the origin of a nonlinear system is uniformly 
asymptotically stable then the linearized system about the origin describes the behavior 
of the nonlinear system.

\section{Criterion for Uniform Contraction}

The question now concerns the criteria to obtain a uniform contraction. There are many 
results in this direction, we suggest Ref. \cite{Coppel}. We present a criterion that best suits 
our purpose. The criterion provides a condition only in terms of the equation, and requires 
no knowledge of the solutions. 

\begin{Teorema}\label{cond:uniCont}
Let $\mathbi{U}(t) = [U_{ij}(t)]$ be a bounded, continuous matrix function on $\mathbb{R}^m$ 
on the half-line and suppose there exists a constant $\eta >0$ such that 
\begin{equation}
U_{ii}(t)  +    \sum_{j=1,  \atop j\not= i}^m |U_{i j }(t)| \le - \eta < 0, 
\label{Uii}  
\end{equation}
for all $t \ge 0$ and $i=1, \cdots , m$. Then the evolution operator is a uniform contraction.
\end{Teorema}

{\it Proof:} We use the norm $ \| \cdot \|_{\infty}$ and its induced norm, see Ex \ref{norInf} in Ap. \ref{LA}. 
Let $\mathbi{ x}(t)$ be a solution. For a fixed time $u >0$ and let $\| \mathbi{x}(u) \|_{\infty}^2 =  x_i (u)^2$.  Note 
$\mathbi{ x}(t)$ is a differentiable function and the norm a continuous function $x_i (t)^2$ will also be the norm in an open 
interval $I = ( u  - a , u +a)$ for some $a>0$.  Therefore, 
\begin{eqnarray}
\frac 1 2 \frac{d}{dt} \| \mathbi{ x}(t) \|_{\infty}^2 &=& \frac 1 2 \frac{d}{dt}  [x_i(t) ]^2 \nonumber \\
&=& x_i(t) \left( \sum_{j=1}^m U_{i j } x_j\right) \nonumber  \\
&=&  U_{ii}(t) x_i^2(t) +  \sum_{j=1, \atop j\not= i}^m U_{i j } x_i(t) x_j(t) \nonumber \\  
&\le& U_{ii}(t) x_i^2(t) +  \sum_{j=1, \atop j\not= i}^m |U_{i j }(t)| x_i^2(t), \nonumber 
\end{eqnarray}
and consequently, 
\begin{eqnarray}
\frac 1 2 \frac{d}{dt} \| \mathbi{x}(t) \|_{\infty}^2 \le \left( U_{ii}(t)  +  \sum_{j=1, \atop j\not= i}^m |U_{i j }(t)| \right)  \| \mathbi{x}(t) \|_{\infty}^2. \nonumber
\end{eqnarray}
Using the condition 
\begin{equation}\label{Uii}  
U_{ii}(t) +    \sum_{j=1, \atop  j\not= i}^m |U_{i j }(t)| \le - \eta < 0,   
\end{equation}
replacing in the inequality 
$$
\frac 1 2  \frac{d}{dt} \| \mathbi{x}(t) \|_{\infty}^2 \le - \eta  \| \mathbi{x}(t) \|_{\infty}^2,
$$
an integration yields
$$
\| \mathbi{x}(t) \|_{\infty}^2 \le \| \mathbi{x} (s) \|_{\infty}^2 - 2 \eta \int_{s}^t \| \mathbi{x}(\tau) \|_{\infty}^2 d \tau, 
$$
for all $t,s \in I$ and $t>s$. Applying the Gronwall inequality we have 
which implies
\begin{equation}\label{boundInfty}
\| \mathbi{x}(t) \|_{\infty} \le e^{- \eta (t-s)} \| \mathbi{x}(s) \|_{\infty}.
\end{equation}
Next note that the argument does not depend on the particular component $i$, because we assume 
that Eq. (\ref{Uii}) is satisfied for any $1 \le  i \le m$. So the norm will satisfy the bound   in Eq. \ref{boundInfty}
for any  compact set of $\mathbb{R}_+$. Noting that all norms are equivalent in finite dimensional 
spaces the result follows $\Box$.

%% file: chapterSync1.tex
\motto{Things which have nothing in common cannot be understood, the one by means of 
the other; the conception of one does not involve the conception of the other \\
\hspace{6cm} --- Spinoza}

\chapter{ Stability of Synchronized Solutions}

We come back to the two fundamental questions concerning the boundedness
of the solutions and the stability of the globally synchronized  in networks
of diffusively coupled oscillators. 

\section{Global Existence of the solutions}

The remarkable property of the networks of diffusively coupled dissipative oscillators is that 
the solutions are always bounded, regardless the coupling strength and network structure. 
The two main ingredients for such boundedness of solutions are:
\begin{description}
\item{ -- Dissipation} of the isolated dynamics given in terms of the Lyapunov function. 
\item{ -- Diffusive} coupling given in terms of the laplacian matrix
\end{description} 
Under these two conditions we can construct a Lyapunov function for the whole
system. The result is then the following

\begin{Teorema} Consider the diffusively coupled network model
\end{Teorema}
$$
\mathbi{x}_i = \mathbi{f}(\mathbi{x}_i) - \alpha \sum_{j=1}^n L_{ij} \mathbi{x}_j,
$$ 
{\it and assume that the isolated system has a 
Lyapunov function satisfying  Assumption \ref{Dissip}. Then, for any network 
the solutions of the coupled equations eventually enter an absorbing domain $\Omega$. 
The absorbing set is independent of the network. }

{\it Proof:} The idea is to construct a Lyapunov function for the coupled 
oscillators in terms of the Lyapunov function of the isolated oscillators. 
Consider the function 
$W : \mathbb{R}^{m n} \rightarrow \mathbb{R}$ where
$$
W(\mathbi{X}) = \frac{1}{2} (\mathbi{X} - \mathbi{A} )^{*} (\mathbi{I}_n \otimes \mathbi{Q} ) ( \mathbi{X} - \mathbi{A})
$$
where $\mathbi{X}$ is given by the vectorization of $(\mathbi{x}_1 , \cdots , \mathbi{x}_n)$ 
and likewise $\mathbi{A} = (\mathbi{1} \otimes \mathbi{ a})^*$, where again $\mathbi{1} = (1, \cdots, 1)$.  
The derivative of the function $W$ along the solutions reads
\begin{eqnarray}
\frac{ d W(\mathbi{X}) }{dt} & =  & (\mathbi{X} - \mathbi{A} )^{*} (\mathbi{I}_n \otimes \mathbi{Q} ) \left[ \mathbi{F}(\mathbi{X}) - 
\alpha \left( \mathbi{L} \otimes \mathbi{I}_m \right) \mathbi{X} \right]  \nonumber \\
& =  & (\mathbi{X} - \mathbi{A} )^{*} (\mathbi{I}_n \otimes \mathbi{Q} ) \mathbi{F}(\mathbi{X}) - \alpha  \mathbi{X}^{*} \left( \mathbi{L} \otimes \mathbi{Q} \right) \mathbi{X} + \alpha \mathbi{A}^{*} \left( \mathbi{L} \otimes \mathbi{Q} \right) \mathbi{ X}, \nonumber 
\end{eqnarray}
however, using the properties of the Kronecker product, see Theorem \ref{Thm:KroPod} and Theorem \ref{Thm:KroT} we have
\begin{eqnarray}
\mathbi{ A}^{*} \left( \mathbi{L} \otimes \mathbi{Q} \right)  &=& ( \mathbi{1} \otimes \mathbi{a})^{*}  \left( \mathbi{L} \otimes \mathbi{Q} \right) \\
& =& \mathbi{1}^* \mathbi{L}  \otimes \mathbi{a}^{*}  \mathbi{Q}  
\end{eqnarray}
but since $\mathbi{1}$  is an eigenvector with eigenvalue $0$ we have $\mathbi{1}^* \mathbi{L} = \mathbi{0}^*$, and consequently 
\begin{eqnarray}
\mathbi{A}^{*} \left( \mathbi{L} \otimes \mathbi{Q} \right) \mathbi{X} = 0.
\end{eqnarray}

Now $\mathbi{L}$ is  positive semi-definite and $\mathbi{ Q}$ is positive definite, hence it follows that $\mathbi{L} \otimes \mathbi{Q}$
is positive semi-definite, see Theorem \ref{Thm:KroP},  and
$$
\mathbi{X}^{*} ( \mathbi{L} \otimes \mathbi{Q})  \mathbi{X} \ge 0.
$$
We have the following upper bound 
\begin{eqnarray}
\frac{ dW  (\mathbi{X})}{dt} &\le& (\mathbi{X} - \mathbi{A} )^{*} (\mathbi{I}_n \otimes \mathbi{Q} ) \mathbi{F}(\mathbi{X})  \nonumber \\
& = & \sum_{i=1}^n (\mathbi{x}_i - \mathbi{a} )^{*} \mathbi{Q} \mathbi{f}(\mathbi{x}_i) \\
& = & \sum_{i=1}^n V^{\prime}(\mathbi{x}_i)
\end{eqnarray}
but by hypothesis $(\mathbi{x}_i - \mathbi{a} )^{*} \mathbi{Q} \mathbi{f}(\mathbi{x}_i)$ is negative on $D \backslash \Omega$, 
hence, $d{W}/dt$ is negative on $D^n \backslash \Omega^n$, since $\Omega$  depends only 
on the isolated dynamics the result follows. $\Box$

This means that the trajectory of each oscillators is bounded
$$
\| \mathbi{x}_i (t) \| \le K
$$
where $K$ is a constant and can be chosen to be independent of the node $i$ and of the 
network parameters such as degree and size.

\section{Trivial example: Autonomous linear equations}

Before we study the stability of the synchronized motion in networks of nonlinear equations,  we address the stability 
problem between two mutually coupled linear equations. The following example is pedagogic and bears all the ideas of the prove 
of the general case. Consider the scalar equation
$$
\frac{d x}{dt} = a x
$$
where $a > 0$. The evolution operator reads
$$
T(t,s) = e^{a(t-s)},
$$
so solutions starting at $x_0$ are given by
$x(t)  = e^{a t}x_0 $. The dynamics is rather simple, for all initial conditions $x_0 \not = 0$
diverge exponentially fast with rate of divergency given by $a$.  Consider two of such equations diffusively coupled
\begin{eqnarray}
\frac{d x_1}{dt} = a x_1 + \alpha ( x_2 - x_1) \nonumber \\
\frac{d x_2}{dt} = a x_2 + \alpha ( x_1 - x_2) \nonumber
\end{eqnarray}

The pain in the neck is that the solutions of the isolated system are not bounded. Since the equation is linear 
the nontrivial solution are not bounded. On the other hand, because the linearity we don't need the boundedness 
of solutions to address synchronization. If $\alpha$ is large enough the two systems will synchronize 
$$
\lim_{t \rightarrow \infty} | x_1(t) - x_2(t) | =0.
$$

Let us introduce 
$$
\mathbi{X} =
\left( 
\begin{array}{c}
x_1 \\
x_2
\end{array}
\right)
$$
The adjacency matrix and Laplacian are given
$$
\mathbi{A} = 
\left( 
\begin{array}{cc}
0 & 1 \\
1 & 0
\end{array}
\right)
\, \mbox{    and    } \, 
\mathbi{L} = 
\left( 
\begin{array}{cc}
1 & -1 \\
-1 & 1
\end{array}
\right)
$$

The coupled equations can be represented as
$$
\frac{d \mathbi{X}}{dt} = \left[  a \mathbi{I}_2 - \alpha \mathbi{L} \right] \mathbi{X}
$$
According to the example \ref{Ex:Linear} the solution reads
\begin{equation}
\mathbi{X}(t) = e^{\left[  a \mathbi{I}_2 - \alpha \mathbi{L} \right]  t  } \mathbi{X}_0.
\label{aL}
\end{equation}

We can compute the eigenvalues and eigenvectors of the Laplacian $\mathbi{L}$. 
An easy computation shows that $\mathbi{1} = (1,1)^*/\sqrt{2}$ is an eigenvector of 
associated with the eigenvalue $0$, and $\mathbi{v}_2 = (1,-1)/\sqrt{2}$ is an eigenvector
associated with the eigenvalue $\lambda_2  = 2$. Note that with respect to the Euclidean inner product the set 
$\{ \mathbi{1}, \mathbi{v}_2 \}$ is an orthonormal basis of $\mathbb{R}^2$.

To solve Eq. (\ref{aL}) we note that if for a given matrix $\mathbi{B}$ we have that $\mathbi{u}$ is an 
eigenvector  associated with  the  eigenvalue $\lambda$.  Then the matrix
$\mathbi{C} = \mathbi{B} - a \mathbi{I}$
has eigenvector $\mathbi{u}$ associated with the eigenvalue $ \lambda - a$.

We can write
$$
\mathbi{X}_0 = c_1 \mathbi{1} + c_2 \mathbi{v}_2,
$$
recalling that $e^{\mathbi{B} t }\mathbi{u} = e^{\lambda t } \mathbi{u}$, and if $\mathbi{B}$ and $\mathbi{C}$ commute 
then $e^{\mathbi{B} + \mathbi{C}} = e^{\mathbi{B}} e^{\mathbi{C}}$. Hence,
the solution of the vector equation $\mathbi{X}(t)$ reads
\begin{eqnarray}
\mathbi{X}(t) &=& e^{\left[  a \mathbi{I}_2 - \alpha \mathbi{L} \right]  t  } \left(  c_1 \mathbi{1} + c_2 \mathbi{v}_2  \right) \\
&=&c_1 e^{a t } \mathbi{1} + c_2 e^{( a - \alpha \lambda_2 ) t } \mathbi{v}_2.
\end{eqnarray}
To achieve synchronization the dynamics along the transversal mode $\mathbi{v}_2$ must be damped out, that is, 
$\lim_{t \rightarrow \infty} c_2 e^{( a - \alpha \lambda_2 ) t } \mathbi{v}_2  = 0$. This implies that  
$$
\alpha \lambda_2 > a \, \, \Rightarrow \alpha > \frac{a}{\lambda_2}
$$

Hence, the coupling strength has to be larger than the rate of divergence of the trajectories 
over the spectral gap. This is a general principle in diffusively networks. 

\section{Two coupled nonlinear equations}
~ 

Let us consider now the stability of two oscillators diffusively coupled. At this time we 
perform the  stability analysis without using the Laplacian properties. This allows a 
simple analysis and provides the condition for synchronization in the same spirit as we shall 
use later on. 

We assume that the nodes are described by Eq. (\ref{nodeq}). 
In the simplest case of two diffusively coupled in all variables systems the 
dynamics is described by
\begin{eqnarray}
\frac{d \mathbi{x}_1}{dt} &=& \mathbi{f}(\mathbi{x}_1) + \alpha (\mathbi{x}_2 - \mathbi{x}_1) \nonumber \\
\frac{d \mathbi{x}_2}{dt} &=& \mathbi{f}(\mathbi{x}_2) + \alpha (\mathbi{x}_1- \mathbi{x}_2) \nonumber
\end{eqnarray}
where $\alpha$ is the coupling parameter. Again, note that 
$$
\mathbi{x}_1(t)=\mathbi{x}_2(t)
$$
defines the synchronization manifold and  is an invariant subspace of the equations 
of motion for all values of the coupling strength. Note that in the subspace the 
coupling term vanishes, and the dynamics is the same as if the systems were uncoupled. 
Hence, we do not control the motion on the synchronization manifold. If the isolated 
oscillators possess a chaotic dynamics, then the synchronized motion will also be 
chaotic.

Again, the problem is then to determine the stability of such subspace in terms of the coupling 
parameter, the coupling strength. It turns out that the subspace it is stable if the coupling 
is strong enough. That is, the two oscillators will synchronize. Note that when they 
synchronize they will preserve the chaotic behavior.

To determine the stability of the synchronization manifold, we analyze the dynamics of the difference
$\mathbi{z} = \mathbi{x}_1 - \mathbi{x}_2$. Our goal is to obtain conditions such that 
$$
\lim_{t \rightarrow \infty } \mathbi{z} = \mathbi{0},
$$
hence, we aim at obtaining the first variational for $\mathbi{z}$.
\begin{eqnarray}
\frac{d \mathbi{z}(t)}{dt } &=& \frac{d \mathbi{x}_1(t)}{dt } - \frac{d \mathbi{x}_2(t)}{dt } \\
& = &  \mathbi{f}(\mathbi{x}_1) - \mathbi{f}(\mathbi{x}_2) -  2 \alpha \mathbi{z} \label{twoC}
\end{eqnarray}
Now if $\| \mathbi{z}(0) \| \ll 1$, we can obtain the first variational equation governing the 
perturbations 
\begin{equation}
\label{var2}
\frac{d \mathbi{z}(t)}{dt } = [  D \mathbi{f}(\mathbi{x}_1(t))  - 2 \alpha \mathbi{I} ] \mathbi{z}.
\end{equation}

The solutions of the variational equation can be written in terms of the 
evolution operator
$$
\mathbi{z}(t) = \mathbi{T}(t,s) \mathbi{z}(s)
$$

Applying Theorem \ref{cond:uniCont}  we obtain conditions for the evolution 
operator to possesses a uniform contraction.  
Let us denote the matrix $D\mathbi{f}(\mathbi{x}_1(t)) = [D\mathbi{f}(\mathbi{x}_1(t))_{ij}]_{i,j=1}^m$.
Uniform contraction requires 
\begin{equation}
\label{Juc}
D\mathbi{f}(\mathbi{x}_1(t))_{ii} - 2 \alpha  + \sum_{j=1, j\not=i}^m | D\mathbi{f}(\mathbi{x}_1(t))_{ij} | < 0 
\end{equation}
for all $t \ge 0$, similarly 
\begin{eqnarray}
\label{auc}
\alpha_c & = & \sup_{\mathbi{x} \in \Omega , \atop 1\le i \le m} \left\{ \sum_{j=1, \atop j\not=i}^m | D\mathbi{f}(\mathbi{x}(t))_{ij} | + D\mathbi{f}(\mathbi{x}(t))_{ii} \right\},  \nonumber \\
\end{eqnarray}
since $\Omega$ is limited and connected in virtue of the Weierstrass Theorem $\alpha_c$ exists. 
Note that $\alpha_c$ is closely related to the norm of the Jacobian $\| D\mathbi{f}(\mathbi{x}) \|_{\infty}$.
Interestingly, $\alpha_c$ can be computed only by 
accessing the absorbing domain and the Jacobian.  Note that this bound for  critical coupling 
is usually larger than needed to observe synchronization. However, this bound is general and independent of the trajectories, and
guarantee a stable and robust synchronized motion.  

The trivial solution $\mathbi{z} \equiv \mathbi{0}$ might be stable before we guarantee that the evolution 
operator is a uniform contraction. In this case, however, we don't guarantee that stability of the trivial 
solutions persists under perturbations. Hence, we cannot guarantee that the nonlinear perturbation coming 
from the Taylor remainder does not destroy the stability.  We avoid tackling this case, since it would bring 
only further technicalities. Note the above $\alpha_c$ synchronization is stable under small perturbations

\begin{ex}
Consider the Lorenz system presented in Sec. \ref{LorenzField}.
\end{ex} 
{\it Then}  
$$
[D\mathbi{f}(\mathbi{x}) - \alpha \mathbi{I}_3]  = 
\left(
\begin{array}{ccc}
- \sigma - \alpha & \sigma  & 0 \\ 
r  - z & - \alpha - 1 & -x  \\ 
y & x & -b  - \alpha 
\end{array}
\right),
$$
{\it noting that the trajectories lie within the absorbing domain $\Omega$ given in Proposition \ref{OmeLor}, 
we have}
$$
| x | \le \sqrt{r} \frac{b}{\sqrt{b-1}}, \,   \, \,  \! |y| \le r \frac{b}{\sqrt{\sigma(b-1)}}, \, \, \mbox{  and   }  |z - r| \le r\left( \frac{b}{\sqrt{\sigma(b-1)}} + 1\right), 
$$
therefore, 
$$
\alpha_c   =     r\left(  1+ \frac{b}{\sqrt{\sigma(b-1)}}\right) + \sqrt{r} \frac{b}{\sqrt{b-1}}  -1 
$$
{\it For  the standard parameters (see Sec. \ref{LorenzField}) we have 
$ \alpha_c  \approx 56.2. $ For the two coupled Lorenz, this provides 
the critical parameter for synchronization }
$$
\alpha \ge  \alpha_c / 2  \approx 28.1
$$

\begin{figure}
\centerline{\hbox{\psfig{file=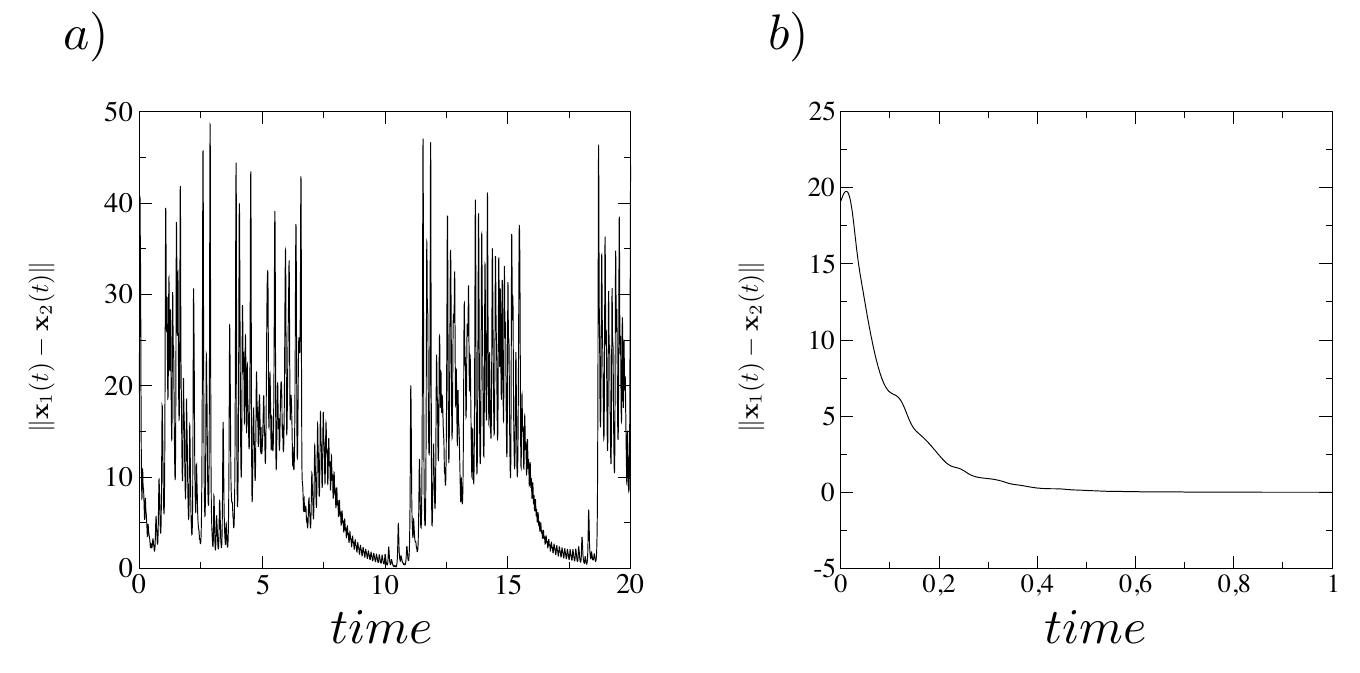,width=12.0cm}}}
\caption{Time evolution of norm  $\| \mathbi{x}_1(t) - \mathbi{x}_2(t) \|$  for distinct initial conditions. $a)$ 
For $\alpha = 0.3$ we observe an asynchronous behavior. $b)$ for  $\alpha = 27$ above the critical coupling 
parameter  the norm of the difference vanishes exponential fast as a function of times, just as predicted by 
the uniform contraction.}
\label{LorenDifNorm}
\end{figure}

We have simulated the dynamics of Eq. (\ref{twoC}) using the Lorenz system. For $\alpha = 27 > \alpha_c$
we observe that the complete synchronized state is stable. If the two Lorenz systems start at distinct
initial condition as time evolves the difference vanishes exponentially fast, see Fig \ref{LorenDifNorm}

\begin{figure}
\centerline{\hbox{\psfig{file=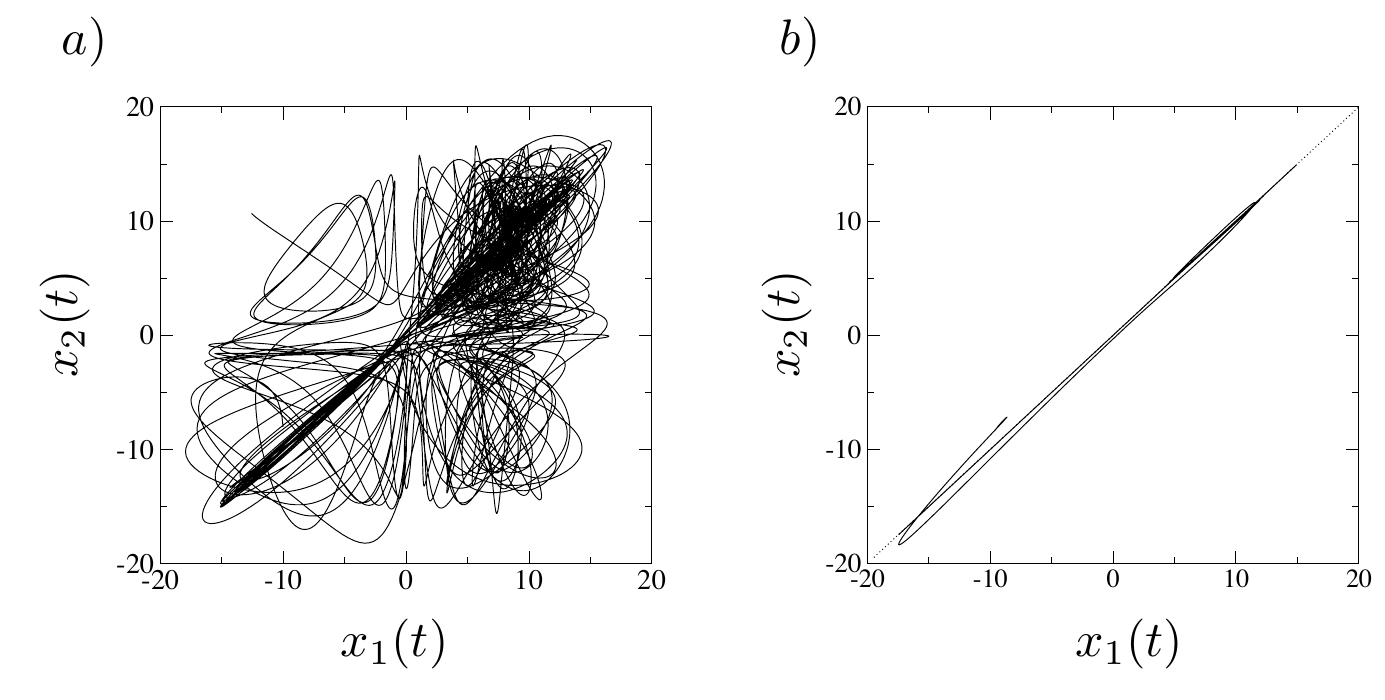,width=12.0cm}}}
\caption{Behavior of the trajectories in the projection $x_1 \times x_2$. $a)$ for the coupling parameter 
$\alpha = 3$, the trajectories are out of sync, and spread around. $b)$ for $\alpha = 27$ the trajectories 
converge to the diagonal line  $x_1 = x_2$. Trajectories in a neighborhood of the diagonal converge to it 
exponentially fast.}
\label{LorenDiag}
\end{figure}

If we depict $x_1 \times x_2$ the dynamics will lie on a diagonal subspace $x = y$. If the initial conditions
start away from the diagonal $x = y$ the evolution time series will then converge to it, see Fig \ref{LorenDiag}

\section{Network Global Synchronization}

We turn to the stability problem in networks. Basically the same conclusion 
as before holds: the network is synchronizable for strong enough coupling strengths. 
In such a case we want to determine the critical coupling in relation to the network structure.  
A positive answer to these question is given by the following

\begin{Teorema} \label{MainThm}Consider the diffusively coupled network model
\end{Teorema}
$$
\mathbi{x}_i = \mathbi{f}(\mathbi{x}_i) + \alpha \sum_{j=1}^n A_{ij} (\mathbi{x}_j - \mathbi{x}_i),
$$ 
{\it on a connected network. Assume that the isolated system has a 
Lyapunov function satisfying  Assumption \ref{Dissip} with an absorbing domain $\Omega$. 
Moreover, assume that  for a given time $s\ge 0$ all trajectories are in a neighborhood 
of the synchronization manifold lying on the absorbing domain 
$\Omega$, and consider the $\alpha_c$ given by Eq. (\ref{auc}), 
and $\lambda_2$ the smallest nonzero eigenvalue of the Laplacian.  Then, for any}  
$$
\alpha > \frac{\alpha_c}{\lambda_2},
$$
{\it the global synchronization is uniformly asymptotically stable. Moreover, the transient to the globally 
synchronized behavior is given the algebraic connectivity, that is, for any $i$ and $j$  
$$
\| \mathbi{x}_i(t) - \mathbi{x}_j(t) \| \le M e^{- (\alpha \lambda_2  - \alpha_c) t}
$$
}
~\\

The above result relates the  threshold coupling  for synchronization in contributions coming  solely 
from dynamics $\alpha_c$, and network structure $\lambda_2$. Therefore, for a fixed node dynamics 
we can analyze how distinct network facilitates or inhibits global synchronization. 
To continue our discussion we need the following
\begin{Definicao}
Let $\beta(G)$ be the critical coupling parameter for the network $G$. 
We say that the network $G$ is better synchronizable than $H$ if for fixed node dynamics
$$
\beta(G) < \beta (H)
$$ 
\end{Definicao}
Recalling the general bounds presented in Theorem \ref{boundl} 
we conclude that the complete network is the most synchronizable network.
Furthermore, the following general statement is also true
\begin{description}
\item{--}  {\it For a fixed network size, network with small diameter are better synchronizable. 
Hence, the ability of the network to synchronize depends on the overall connectedness of the
graph. }
\end{description}

Recall the results presented in table  \ref{table}, and let denote $\alpha_c$ denote the 
critical coupling parameter, the dependence of $\alpha_c$ in terms of the network size can be seen table \ref{tableLambda} 
\begin{table}
\label{tableLambda}
\caption{Leading order dependence of $\beta$ on the network size for the networks in Fig. \ref{Gexam}}
\begin{center}
\begin{tabular}{ccc}
\hline
Network 	&	\hspace{0.5cm} & $\beta $   		 \\
\hline
\hline
~\\
Complete	&	~ &  $\displaystyle \frac{1}{n} $ \\
~\\
ring	&	~ & $ \displaystyle \frac{n^2}{2}$  \\
\\
Star & ~ & $\displaystyle 1$  \\
\\
\hline
\end{tabular}
\end{center}
\end{table}

The difficulty to synchronize a complete network decreases with the network size, 
whereas to synchronize the cycle increases quadratically with the size. 

Now we present the proof of Theorem \ref{MainThm}. We omit some details that are not relevant for the understanding of the proof. A full discussion of the proof can be found in \cite{Jaap}We must show that the synchronization 
manifold $\mathcal{M}$ is locally attractive. In other words, whenever the nodes start close 
together they tend to the same future dynamics, that is, $\| \mathbi{x}_i (t)  - \mathbi{x}_j (t) \| \rightarrow 0$, 
for any $i$ and $j$. For pedagogical purposes we split the proof into four main steps.

\mathbi{ Step 1: Expansion into  the Laplacian Eigenmodes.} Consider the equations of motion 
in the block form 
$$
\frac{d \mathbi{X}}{dt } = \mathbi{F}(\mathbi{X}) - \alpha (\mathbi{L} \otimes \mathbi{I}_m ) \mathbi{X}
$$
Note that since $L$ is symmetric, by Theorem \ref{LA:ThmDiag} there exists an  orthogonal matrix  $\mathbi{O}$ such that 
$$
\mathbi{L} = \mathbi{O} \mathbi{ M} \mathbi{O}^{*},
$$
where $\mathbi{M} = $ diag$(\lambda_0, \lambda_1, \ldots, \lambda_n)$  is the eigenvalue matrix. 
Introducing
$$
\mathbi{Y} = \mbox{col}(\mathbi{y}_1,\mathbi{y}_2, \cdots,\mathbi{y}_n)
$$
we can write the above equation in terms of Laplacian eigenvectors 
\begin{eqnarray}
\mathbi{X} &=& \left( \mathbi{O}  \otimes \mathbi{I}_m \right) \mathbi{Y}, \nonumber \\
& = & \sum_{i=1}^n \mathbi{v}_i \otimes \mathbi{y}_i \nonumber
\end{eqnarray}

For sake of simplicity we call $\mathbi{y}_1 = \mathbi{s}$, and remember that now note that 
$\mathbi{v}_1 = \mathbi{1}$ hence
$$
\mathbi{X} = \mathbi{1} \otimes \mathbi{s} + \mathbi{U},
$$
where 
$$
\mathbi{U} = \sum_{i=2}^n \mathbi{v}_i \otimes \mathbi{y}_i.
$$
In this way we split the contribution in the direction of the global synchronization 
and \mathbi{U}, which accounts for the contribution of the transversal. Note that if 
$\mathbi{U}$ converges to zero then the system completely synchronize, that is 
$\mathbi{X}$ converges to $\mathbi{1} \otimes \mathbi{s}$ which clearly implies that
$$
\mathbi{x}_1 = \cdots = \mathbi{x}_n = \mathbi{s}
$$

The goal then is to obtain conditions so that $\mathbi{U}$ converges to zero.
\\

\mathbi{ Step 2: Variational equations for the Transversal Modes.} The equation of motion in terms of the 
Laplacian modes decomposition reads
\begin{eqnarray}
\frac{d \mathbi{X}}{dt } &=& \mathbi{F}(\mathbi{X}) -\alpha(\mathbi{L} \otimes \mathbi{I})  \mathbi{X}  \nonumber \\
\mathbi{1} \otimes \frac{d \mathbi{s}}{dt } + \frac{d \mathbi{U}}{dt }&=& \mathbi{F} (\mathbi{1} \otimes \mathbi{s} + \mathbi{U})  - \alpha(\mathbi{L} \otimes \mathbi{I})\left(  \mathbi{1} \otimes \mathbi{s} + \mathbi{U}\right), \nonumber
\end{eqnarray} 
We assume that $\mathbi{U}$ is small and perform a Taylor expansion about the synchronization manifold. 
$$
\mathbi{F} (\mathbi{1} \otimes \mathbi{s} + \mathbi{U})  = \mathbi{F} (\mathbi{1} \otimes \mathbi{s})  + D \mathbi{F} (\mathbi{1} \otimes \mathbi{s}) \mathbi{U}  + \mathbi{R}(\mathbi{U}), 
$$
where $\mathbi{R}(\mathbi{U})$ is the Taylor remainder $\| \mathbi{R} (\mathbi{U}) \| = O(\| \mathbi{U} \|^2) $.
Using the Kronecker product properties \ref{Thm:KroPod} and  the fact that $\mathbi{L} \mathbi{1} = \mathbi{0}$, together with 
$$
\mathbi{1} \otimes \frac{d \mathbi{s}}{dt } = \mathbi{F} (\mathbi{1} \otimes \mathbi{s}) = \mathbi{1} \otimes \mathbi{f}(\mathbi{s})
$$
and likewise
$$
 D \mathbi{F} (\mathbi{1} \otimes \mathbi{s}) \mathbi{U}  = [  \mathbi{I}_n \otimes D \mathbi{f} (\mathbi{s}) ]  \mathbi{U}, 
$$
and we have
\begin{eqnarray}
\frac{d \mathbi{U}}{dt }= [ D \mathbi{F} (\mathbi{1} \otimes \mathbi{s})  - \alpha (\mathbi{L} \otimes \mathbi{I}) ]   \mathbi{U} + \mathbi{R}(\mathbi{U}),
\end{eqnarray} 
Therefore, the first variational equation for the transversal modes reads
$$
\frac{d \mathbi{U}}{dt } = [\mathbi{I}_n \otimes D \mathbi{f} (\mathbi{s}) - \alpha \mathbi{L} \otimes \mathbi{I}_m] \mathbi{U}. 
$$

The solution of the above equation has a representation in terms of the evolution operator 
$$
\mathbi{U}(t) = \mathbi{T}(t,s) \mathbi{U}(s)
$$

We want to obtain conditions for the trivial solution of the above to be uniformly asymptotically stable, 
that is, so that the evolution operator is a uniform contraction.

\mathbi{ Step 3: Stabilization of the Transversal  Modes.} Instead of analyzing the full set of equations, 
we can do much better by projecting the equation  into the transversal modes $\mathbi{v}_i$.
Applying $\mathbi{v}_j^* \otimes \mathbi{I}_m$ on the right in the equation for $\mathbi{U}$, it yields
\begin{eqnarray}
\mathbi{v}_j^* \otimes \mathbi{I}_m \left( \sum_{i=2}^n \mathbi{v}_i \otimes \frac{ d \mathbi{y}_i}{dt} \right)  &=&  \mathbi{v}_j^* \otimes \mathbi{I}_m
\left( \sum_{i=2}^n  \mathbi{v}_i \otimes D \mathbi{f} (\mathbi{s}) \mathbi{y}_i  - \alpha \lambda_i \mathbi{v}_i \otimes \mathbi{y}_i \right)  \nonumber \\
\sum_{i=2}^n \mathbi{v}_j^* \mathbi{v}_i \otimes \frac{ d {\mathbi{y}}_i}{dt}  &=&  
\sum_{i=2}^n \mathbi{v}_j^* \mathbi{v}_i \otimes [ D \mathbi{f} (\mathbi{s}) - \alpha \lambda_i  \mathbi{I}_m] \mathbi{y}_i \nonumber 
\end{eqnarray}

But since $\mathbi{v}_i$ form an orthonormal basis we have $$\mathbi{v}_j^* \mathbi{v}_i = \delta_{ij},$$ 
where is $\delta_{ij}$ the Kronecker delta. Hence, we obtain the equation for the coefficients 
$$
\frac{d {\mathbi{y}}_i}{dt}  =   [ D \mathbi{f} (\mathbi{s}) - \alpha \lambda_i  \mathbi{I}_m ] \mathbi{y}_i
$$

All blocks have the same form which are different only by $\lambda_i$, the $i$th eigenvalue of  $L$. 
We can write all the blocks in a parametric form 
\begin{equation}
\label{para}
\frac{d {\mathbi{u}}}{dt} = \mathbi{K}(t)  \mathbi{u},
\end{equation}
where 
$$
\mathbi{K}(t) = D\mathbi{f}(\mathbi{s}(t)) - \kappa \mathbi{I}_m
$$
with $\kappa \in \mathbb{R}$. Hence if $\kappa = \alpha \lambda_i$ we have the equation for the $i$th block. 
This is just the same type of equation we encounter before in the example of the two coupled oscillators,  see Eq. (\ref{var2}).  

Now obtain conditions for the evolution operator of Eq. (\ref{para}) to possess a uniform contraction. This is 
done applying the same arguments discussed in Eqs. \ref{Juc} and \ref{auc}. Therefore, the $i$th block has 
a uniform contraction if $\alpha \lambda_i > \alpha_c$. Now since the spectrum of the Laplacian is ordered, the 
condition for all blocks to be uniformly asymptotically stable is 
$$
\alpha  > \frac{\alpha_c}{\lambda_2}
$$
which yields a critical coupling value in terms of $\alpha_c$ and $\lambda_2$.

Taking $\alpha$ larger than the critical value we have that all blocks have uniform contractions. Let $\mathbi{T}_i(t,s)$ be the 
evolution operator of the $i$th block. Then
\begin{eqnarray}
\|  \mathbi{ y}_i (t) \| &\le&   \| \mathbi{T}_i(t,s)  \mathbi{ y}_i(s)\| \nonumber  \\
&\le& \| \mathbi{T}_i(t,s) \| \|  \mathbi{ y}_i(s)\|, \nonumber \\
\end{eqnarray}
by applying Theorem \ref{cond:uniCont} we obtain
\begin{eqnarray}
\|  \mathbi{ y}_i (t) \| \le K_i e^{-\eta_i (t-s)} \| \mathbi{ y}_i(s)\| \label{by}, \nonumber
\end{eqnarray}
where $\eta_i =  \alpha \lambda_i - \alpha_c$.

\mathbi{ Step 4: Norm Estimates.} Using the bounds for the blocks it is easy to obtain a 
bound for the norm of the evolution operator. Indeed, note that 
\begin{eqnarray}
\|  \mathbi{U}  \|_2 &=& \left\| \sum_{i=2}^n \mathbi{v}_i \otimes \mathbi{y}_i \right\|_2 \nonumber \\
&\le& \sum_{i=2}^n \| \mathbi{v}_i \| \| \mathbi{y}_i \|_2 \nonumber 
\end{eqnarray}
where we have used Theorem \ref{normK} (see Ap. \ref{LA}), therefore, 
\begin{eqnarray}
\|  \mathbi{U}  \|_2 & \le& \sum_{i=2}^n \| \mathbi{v}_i \| K_i  e^{-\eta_i (t-s)} \| \mathbi{ y}_i(s)\| \nonumber
\end{eqnarray}

Now using that $e^{-\eta_i (t-s)} \le e^{ - (\alpha \lambda_2 - \alpha_c)}$, and applying 
Theorem \ref{UniCon} we obtain
$$
\|  \mathbi{T}(t,s) \|_2 \le  M e^{ - \eta (t-s)}
$$
with $\eta  = \alpha \lambda_2 - \alpha_c$ for any $t\ge s$. 

By the principle of linearization Lemma \ref{PL}, we conclude that the Taylor 
remainder does not affect the stability of the trivial solution, which correspond to the 
global synchronization. 

The claim about the transient is straightforward, indeed note that 
$$
\| \mathbi{X}(t) - \mathbi{1} \otimes \mathbi{s}(t) \| \le M e^{ - \eta (t-s)} \| \mathbi{U}(s) \|
$$
implying that $\| \mathbi{x}_i(t) - \mathbi{s}(t) \| \le K e^{-\eta(t-s)}$ and 
$$
\|  \mathbi{x}_i(t) -  \mathbi{x}_j(t) \| \le \|  \mathbi{x}_i(t) -  \mathbi{s}(t) \| + \|  \mathbi{x}_i(t) -  \mathbi{s}(t) \| 
$$
in virtue of the triangular triangular inequality, and we concluding the proof.  $\Box$

%% file: chapterClusterHyper.tex
\motto{
What seems obvious is often only obvious after it is understood. \\
\hspace{3cm} -- Abel}

\chapter{Some Generalizations}

\section{Cluster Synchronization}

Let $\mathbi{P} \in \mathbb{R}^{n \times n}$ be a permutation matrix that encodes a symmetry of the network. If
\begin{equation}
\mathbi{P}\mathbi{A} = \mathbi{A}\mathbi{P},
\end{equation}
Then the adjacency structure is invariant under the permutation of node labels defined by $\mathbi{P}$. Since the degree matrix $\mathbi{D}$ satisfies $\mathbi{P}\mathbi{D} = \mathbi{D}\mathbi{P}$, we also have:
\begin{equation}
\mathbi{P}\mathbi{L} = \mathbi{L}\mathbi{P}.
\end{equation}
Lets consider some examples

{\bf Three-node path}

\[
\mathbi{A} = \begin{bmatrix}
0 & 1 & 0 \\
1 & 0 & 1 \\
0 & 1 & 0
\end{bmatrix}, \quad
\mathbi{P} = \begin{bmatrix}
0 & 0 & 1 \\
0 & 1 & 0 \\
1 & 0 & 0
\end{bmatrix}
\]

Here, \( \mathbi{P} \) swaps nodes 1 and 3. Then:
\[
\mathbi{AP} = \mathbi{P A} = \mathbi{A}
\]

{\bf Four-node Example }

Let the adjacency matrix be and two permutations:
\[
\mathbi{A} = \begin{bmatrix}
0 & 1 & 1 & 0 \\
1 & 0 & 0 & 1 \\
1 & 0 & 0 & 1 \\
0 & 1 & 1 & 0
\end{bmatrix}, \qquad
\mathbi{P}_{14} = \begin{bmatrix}
0 & 0 & 0 & 1 \\
0 & 1 & 0 & 0 \\
0 & 0 & 1 & 0 \\
1 & 0 & 0 & 0
\end{bmatrix}
\qquad
\mathbi{P}_{23} = \begin{bmatrix}
1 & 0 & 0 & 0 \\
0 & 0 & 1 & 0 \\
0 & 1 & 0 & 0 \\
0 & 0 & 0 & 1
\end{bmatrix}
\]
swapping nodes $1$ and $4$ then $2$ and $3$. One can verify:
\[
\mathbi{AP} = \mathbi{PA} = \begin{bmatrix}
0 & 1 & 1 & 0 \\
1 & 0 & 0 & 1 \\
1 & 0 & 0 & 1 \\
0 & 1 & 1 & 0
\end{bmatrix} = \mathbi{A}
\]
where $\mathbi{P}$ is either $\mathbi{P}_{14}$ or $\mathbi{P}_{23}$

\subsection{Cluster Synchronization Manifold}

Let
\[
\mathbi{Q} := \mathbi{I} - \mathbi{P},
\]
where $\mathbi{P}$ is a permutation matrix. Define the cluster synchronization manifold as
\[
\mathcal{C} := \ker(\mathbi{Q} \otimes \mathbi{I}_m ) \subset \mathbb{R}^{n m}.
\]

\noindent
We now show that this definition is equivalent to the fixed-point subspace of the group action induced by the permutation:
\[
\mathcal{C} = \mathrm{Fix}(\mathbi{P} \otimes \mathbi{I}_m) := \{ \mathbi{x} \in \mathbb{R}^{nm} : (\mathbi{P} \otimes \mathbi{I}_m)\mathbi{x} = \mathbi{x} \}.
\]

\noindent
Indeed, note that
\[
\mathbi{x} \in \ker(\mathbi{Q} \otimes \mathbi{I}_m) \iff (\mathbi{Q} \otimes \mathbi{I}_m)\mathbi{x} = \mathbi{0} \iff (\mathbi{I} - \mathbi{P}) \otimes \mathbi{I}_m \cdot \mathbi{x} = 0 \iff (\mathbi{P} \otimes \mathbi{I}_m)\mathbi{x} = \mathbi{x}.
\]
Therefore:
\[
\ker(\mathbi{Q} \otimes \mathbi{I}_m) = \mathrm{Fix}(\mathbi{P} \otimes \mathbi{I}_m).
\]

\noindent
This space consists of all vectors in which states of nodes mapped to each other by \( \mathbi{P} \) are identical. \( \mathcal{C} \) captures the cluster synchronization pattern encoded by the permutation symmetry.

\subsection{Invariance of the Cluster Synchronization Manifold}

Let $\mathbi{x} \in \mathcal{C}$ so that $(\mathbi{P} \otimes \mathbi{I}_m)\mathbi{x} = \mathbi{x}$. Then:
\begin{align*}
\dot{\mathbi{x}} &= \mathbi{F}(\mathbi{x}) - \alpha (\mathbi{L} \otimes \mathbi{H})\mathbi{x}, \\
(\mathbi{P} \otimes \mathbi{I}_m)\dot{\mathbi{x}} &= (\mathbi{P} \otimes \mathbi{I}_m)\mathbi{F}(\mathbi{x}) - \alpha (\mathbi{P} \otimes \mathbi{I}_m)(\mathbi{L} \otimes \mathbi{H})\mathbi{x}.
\end{align*}
Since $\mathbi{P}$ permutes nodes and $\mathbi{f}$ is the same at all nodes, we have:
\[
(\mathbi{P} \otimes \mathbi{I}_m)\mathbi{F}(\mathbi{x}) = \mathbi{F}((\mathbi{P} \otimes \mathbi{I}_m)\mathbi{x}) = \mathbi{F}(\mathbi{x}),
\]
and from $\mathbi{PL} = \mathbi{LP}$, it follows that:
\[
(\mathbi{P} \otimes \mathbi{I}_m)(\mathbi{L} \otimes \mathbi{H}) = (\mathbi{L} \otimes \mathbi{H})(\mathbi{P} \otimes \mathbi{I}_m).
\]
Hence,
\[
(\mathbi{P} \otimes \mathbi{I}_m)\dot{\mathbi{x}} = \dot{\mathbi{x}},
\]
which implies that $\dot{\mathbi{x}} \in \ker(\mathbi{Q} \otimes \mathbi{I}_m)$ whenever $\mathbi{x} \in \ker(\mathbi{Q} \otimes \mathbi{I}_m)$. Therefore, the manifold $\mathcal{C}$ is invariant under the flow.

\subsection{Spectrum of \emph{\( \mathbi{LQ} \)}}

The commutation relation $\mathbi{PA} = \mathbi{AP}$ implies that $\mathbi{A}$ preserves each eigenspace of $\mathbi{P}$. Therefore, $\mathbi{A}$ and $\mathbi{P}$ admit a common orthonormal eigenbasis. That is, there exist vectors $\mathbi{v}_j$ such that
\[
    \mathbi{A} \mathbi{v}_j = \lambda_j \mathbi{v}_j,
    \qquad
    \mathbi{P} v_j = {p}_j \mathbi{v}_j,
\]
where $\lambda_j \in \sigma(\mathbi{A})$ and $p_j \in \sigma(\mathbi{P})$ (in particular $|p_j| = 1$). Moreover, 
\[
    \mathbi{Q} \mathbi{v}_j = (1 - p_j) \mathbi{v}_j.
\]

Applying $\mathbi{AP}$ to $\mathbi{v}_j$ yields
\[
    (\mathbi{AP})\mathbi{v}_j 
    = \mathbi{A}(\mathbi{P}v_j)
    = \mathbi{A}(p_j \mathbi{v}_j)
    = p_j \mathbi{A} \mathbi{v}_j
    = p_j \lambda_j \mathbi{v}_j.
\]
Hence $\mathbi{v}_j$ is an eigenvector of $\mathbi{AP}$ with eigenvalue $p_j \lambda_j$, so
\[
    \sigma(\mathbi{AP})
    =
    \{\, p_j \lambda_j : \lambda_j \in \sigma(\mathbi{A}),\; p_j \in \sigma(\mathbi{P}) \,\}.
\]

Notice that  $\mathbi{P}$ and $\mathbi{A}$ commute, each eigenspace of $\mathbi{P}$ is $\mathbi{A}$-invariant.  Therefore, we may choose a basis $\{\mathbi{v}_j\}$ consisting of \emph{common} eigenvectors of both $\mathbi{P}$ and $\mathbi{A}$.
The eigenvalues $\lambda_j \in \sigma(\mathbi{A})$ and $p_j \in \sigma(\mathbi{P})$ are thus associated with the same eigenvector $\mathbi{v}_j$, 
which uniquely determines their pairing in the product $p_j \lambda_j$. The ordering is therefore determined by the 
choice of eigenvector, not independently.

Notice that $\mathbi{L}$ and $\mathbi{P}$ can also be simultaneously diagonalized in the same basis $\{\mathbi{v}_j\}$.
If $\mathbi{L} v_j = \mu_j \mathbi{v}_j$, then using $\mathbi{Q} \mathbi{v}_j = (1 - p_j) \mathbi{v}_j$, we compute
\[
    (\mathbi{QL})\mathbi{v}_j = \mu_j (1 - p_j) \mathbi{v}_j.
\]
Therefore
\[
    \sigma(\mathbi{QL})
    =
    \{\, (1 - p_j)\,\mu_j : \mu_j \in \sigma(\mathbi{L}),\; p_j \in \sigma(\mathbi{P}) \,\}.
\]

\medskip
\noindent
In particular, the synchrony subspace is
\[
\operatorname{Fix}(\mathbi{P}) = \{\, \mathbi{x} \in \mathbb{R}^n : \mathbi{P}\mathbi{x} = \mathbi{x} \,\}.
\]
Since $\mathbi{P}v_j = p_j \mathbi{v}_j$ for eigenvectors $\mathbi{v}_j$ of $\mathbi{P}$, we have
\[
\mathbi{v}_j \in \operatorname{Fix}(\mathbi{P})
\quad\Longleftrightarrow\quad
p_j = 1.
\]
Therefore the synchrony space is precisely the eigenspace of $\mathbi{P}$ associated
with the eigenvalue $p_j = 1$, while all transverse directions satisfy $p_j \neq 1$.
For such $p_j's$, the associated eigenvalue of $\mathbi{QL}$ vanishes, which is consistent with the fact that $\mathbi{QL}$ acts trivially on the synchronization manifold $\ker(\mathbi{Q})$.

Since $[\mathbi{L}, \mathbi{Q}] = 0$, we can simultaneously diagonalize $\mathbi{L}$ and $\mathbi{Q}$. Suppose the nodes are ordered such that the synchronized clusters come first. Then, in an orthonormal basis adapted to $\ker \mathbi{Q}$ and $\mathrm{Im} \mathbi{Q}$, the matrix $\mathbi{LQ}$ takes the block form:
\[
\mathbi{T}^\top (\mathbi{LQ}) \mathbi{T} = 
\begin{bmatrix}
0 & 0 \\
0 & \mathbi{L}_\perp \mathbi{Q}_\perp
\end{bmatrix},
\]
where $\mathbi{T}$ is an orthogonal matrix whose columns are eigenvectors of $\mathbi{Q}$ and $\mathbi{L}$, and $\mathbi{L}_\perp \mathbi{Q}_\perp$ acts on the transverse (non-synchronized) subspace. Thus,
\[
\sigma(\mathbi{LQ}) = \{0\}^{\dim \ker \mathbi{Q}} \cup \sigma(\mathbi{L}_\perp \mathbi{Q}_\perp),
\]
where the nonzero part determines the dynamics transverse to the manifold.

\begin{theorem}
Let \( \mathbi{P} \in \mathbb{R}^{n \times n} \) be a permutation matrix and define \( \mathbi{Q} := \mathbi{I}_n - \mathbi{P} \). Then the restriction of \( \mathbi{Q} \) to the orthogonal complement of \( \ker \mathbi{Q} \),
\[
\mathbi{Q}_\perp := \mathbi{Q}|_{\mathrm{Fix}(\mathbi{P})^\perp},
\]
has no zero eigenvalues. In other words,
\[
\sigma(\mathbi{Q}_\perp) \subset \mathbb{C} \setminus \{0\}.
\]
\end{theorem}

\begin{proof}
Since \( \mathbi{P} \) is a real permutation matrix, it is orthogonal and hence unitarily diagonalizable. Its eigenvalues lie on the unit circle in the complex plane, i.e., 
\[
\sigma(\mathbi{P}) \subset \{ e^{2\pi i k / m} : k \in \mathbb{Z} \}.
\]

Let \( \{\mathbi{v}_1, \dots, \mathbi{v}_n\} \) be an orthonormal basis of eigenvectors of \( \mathbi{P} \), with \( \mathbi{P}\mathbi{v}_j = \pi_j \mathbi{v}_j \) for some \( \pi_j \in \mathbb{C} \), \( |\pi_j| = 1 \). Then:
\[
\mathbi{Q} \mathbi{v}_j = (\mathbi{I} - \mathbi{P})\mathbi{v}_j = (1 - \pi_j)\mathbi{v}_j.
\]

Thus, the eigenvalues of \( \mathbi{Q} \) are \( \{1 - \pi_j\} \). In particular, if \( \pi_j = 1 \), then \( \mathbi{Q} \mathbi{v}_j = 0 \), and \( \mathbi{v}_j \in \ker \mathbi{Q} = \mathrm{Fix}(\mathbi{P}) \). If \( \pi_j \ne 1 \), then \( \mathbi{Q} \mathbi{v}_j = (1 - \pi_j)\mathbi{v}_j \ne 0 \), and \( \mathbi{v}_j \in \mathrm{Fix}(\mathbi{P})^\perp \).
Therefore, the eigenvalues of the restriction \( \mathbi{Q}_\perp \) are \( \{1 - \pi_j : \pi_j \ne 1\} \), all of which are nonzero. Hence,
\[
\sigma(\mathbi{Q}_\perp) \subset \mathbb{C} \setminus \{0\}.
\]
\end{proof}

\subsection{Transverse Stability of the Cluster Synchronization Manifold}

Linearizing the dynamics around a solution $\mathbi{s}(t) \in \mathcal{C}$, we write:
\[
 \dot{\mathbi{z}} = D\mathbi{F}(\mathbi{s}(t)) \mathbi{z} - \alpha (\mathbi{L} \otimes \mathbi{H})  \mathbi{z}.
\]
Decomposing $ \mathbi{x}$ into components in $\ker(\mathbi{Q} \otimes \mathbi{I}_m)$ and its complement $\mathrm{Im}(\mathbi{Q} \otimes \mathbi{I}_m)$, and projecting onto the transverse subspace, we obtain:
\[
 \dot{\mathbi{y}} = \left[ D\mathbi{f}(\mathbi{s}(t)) - \alpha (\mathbi{L}_\perp \mathbi{Q}_\perp) \otimes \mathbi{H} \right]  \mathbi{y}.
\]
The cluster synchronization manifold $\mathcal{C}$ if the synchronization criterion we established before is met.

\subsection{Examples}

{\bf Three–node example}.
Recall that the order of multiplication between the spectrum of $\mathbf{L}$ and $\mathbf{Q}$ is determined by the eigenspaces. Notice that the synchrony subspace is
\[
\operatorname{Fix}(\mathbi{P})
= \{\, \mathbi{x} \in \mathbb{R}^3 : x_1 = x_3 \, \}
= \operatorname{span}\left\{ (1,0,1)^\top,\; (0,1,0)^\top \right\}.
\]
A transverse direction (orthogonal to $\operatorname{Fix}(\mathbi{P})$) is given by
\[
\mathbi{v}_\perp = (-1,0,1)^\top.
\]
In this case, $\mathbi{L}_{\perp}\mathbi{Q}_{\perp}\mathbi{v}_\perp = 2 \mathbi{v}_\perp$.
The spectrum is
\[
\sigma(\mathbi{L})=\{0,1,3\}, \quad \sigma(\mathbi{P})=\{1,1,-1\}, \quad \sigma(\mathbi{Q})=\{0,0,2\} \quad \Rightarrow \quad
\sigma(\mathbi{L}_{\perp}\mathbi{Q}_{\perp}) = \{2\}.
\]


\noindent
{\bf The four–node example} Here we have two clusters $\{1,4\}$ and $\{2,3\}$. We will present the analysis for the cluster $\{1,4\}$. The analysis for the other cluster is similar.  
The permutation $P_{14}$ swaps nodes $1$ and $4$:
\[
\mathbi{P}_{14} =
\begin{bmatrix}
0 & 0 & 0 & 1 \\
0 & 1 & 0 & 0 \\
0 & 0 & 1 & 0 \\
1 & 0 & 0 & 0
\end{bmatrix}.
\]
Its fixed--point subspace is
\[
\operatorname{Fix}(\mathbi{P}_{14})
= \{ \mathbi{x} \in \mathbb{R}^4 : \mathbi{P}_{14}\mathbi{x} = \mathbi{x} \}= \{ (a,b,c,a)^\top : a,b,c \in \mathbb{R} \}.
\]
A convenient basis is
\[
\operatorname{Fix}(\mathbi{P}_{14}) = {\rm span}\bigl\{ (1,0,0,1)^\top,\; (0,1,0,0)^\top,\; (0,0,1,0)^\top \bigr\}.
\]

The orthogonal complement  is
\begin{eqnarray}
\operatorname{Fix}(\mathbi{P}_{14})^\perp &=& \{ \mathbi{v} \in \mathbb{R}^4 : \mathbi{v} \cdot \mathbi{w} = 0 \ \text{for all } \mathbi{w} \in \operatorname{Fix}(\mathbi{P}_{14}) \} \nonumber \\
&=& \operatorname{span}\{ (1,0,0,-1)^\top \}.\nonumber
\end{eqnarray}
The vector $(1,0,0,-1)^\top$ represents a perturbation that \emph{breaks} synchrony inside the cluster $\{1,4\}$. 
Moreover, notice that 
\[
\mathbi{L}\mathbi{v}_\perp = 2\mathbi{v}_\perp,
\qquad \mathbi{Q}_{14}\mathbi{v}_\perp = 2\mathbi{v}_\perp \qquad\Rightarrow \qquad
\mathbi{L Q}_{14} \mathbi{v}_\perp = 4 \mathbi{v}_\perp.
\]
implying that 
$$
\sigma(\mathbi{L}_\perp (\mathbi{Q}_{14})_\perp) = \{4\}
$$

{\bf Stability of the Cluster.}
Since the synchronization criterion can be equally applied for both cluster synchronization and global synchronization, we obtain
\begin{eqnarray}
\alpha_{\min}^{\rm global} &:=& \frac{\alpha_c}{\lambda_2(\mathbi{L})} \mbox{~for global sync} \nonumber \\
\alpha_{\min}^{\rm cluster} &:=& \frac{\alpha_c}{\lambda_{\min}(\mathbi{L}_{\perp} Q_{\perp})} \mbox{~for cluster sync}  \nonumber
\end{eqnarray}

\noindent
For these examples, it takes twice less coupling to get cluster synchronization
$$
\alpha_{\min}^{\rm cluster} = \frac{1}{2} \alpha_{\min}^{\rm global}
$$
This means that increase the coupling parameter $\alpha$ we first see cluster synchronization and further increase then a full synchronization.

\section{Hypernetworks}

Beyond pairwise coupling, many real networks interact through triples such as chemical reactions \cite{EddieNatCom}, social influence \cite{Bick}. These hyperorder interactions exhibit interesting dynamics \cite{EddieHyper,CoheRalf}. We will discuss the simplest case.

Consider the equations of the following form
\begin{equation}\label{eq:model}
\mathbi{x}_i' \;=\; \mathbi{f}(\mathbi{x}_i)\;+\;\alpha \sum_{j,k=1}^N A_{ijk}\, \mathbi{H}\!\big(\mathbi{x}_j - 2\mathbi{x}_i + \mathbi{x}_k\big),
\end{equation}
where $\mathbi{x}_i\in\mathbb{R}^m$, $\mathbi{f}:\mathbb{R}^m\to\mathbb{R}^m$, $\mathbi{H}:\mathbb{R}^m\to\mathbb{R}^m$ is the coupling function will be the identity, $\alpha\in\mathbb{R}$, and $A_{ijk}\ge 0$ are weights encoding triplets. Define the two \emph{projected} weighted graphs by
\begin{equation}\label{eq:B1B2}
B^{(1)}_{ij}:=\sum_{k=1}^N A_{ijk}, \qquad
B^{(2)}_{ik}:=\sum_{j=1}^N A_{ijk}.
\end{equation}

These projections have an intuitive meaning. It counts the number of higher-order interactions in which each pair $(i,j)$ participates, along with any third node. That is, 
$$
\sum_{k=1}^N A_{ijk} = \mbox{Number of hyperedges containing both} i  \mbox{and} j
$$
The projection yields an effective pairwise adjacency that is consistent with the hypernetwork structure.
When $A_{ijk}$ is symmetric under swapping $j\leftrightarrow k$ , we have $\mathbi{B}^{(1)}=\mathbi{B}^{(2)}$. If hyperedges are oriented then generally $B^{(1)}\neq B^{(2)}$.
%

\subsection{Decomposition}

\begin{proposition}[Exact decomposition for linear $H$]\label{prop:linearH}
Suppose $H$ is linear. Then for each $i$,
\emph{
\[
\sum_{j,k} A_{ijk}\, \mathbi{H}(\mathbi{x}_j - 2\mathbi{x}_i + \mathbi{x}_k)
\;=\; \sum_j B^{(1)}_{ij}\, \mathbi{H}(\mathbi{x}_j - \mathbi{x}_i)\;+\; \sum_k B^{(2)}_{ik}\, \mathbi{H}(\mathbi{x}_k - \mathbi{x}_i).
\]
}
\end{proposition}

\begin{proof}
Since $\mathbi{x}_j - 2\mathbi{x}_i + \mathbi{x}_k = (\mathbi{x}_j-\mathbi{x}_i)+(\mathbi{x}_k-\mathbi{x}_i)$ and $\mathbi{H}$ is linear. Summing over $j,k$ yields the stated identity, and collecting by pairs $(i,j)$ and $(i,k)$ gives the two projected graphs \eqref{eq:B1B2}. 
\end{proof}

\begin{proposition}[Variational Equation]\label{prop:variational}
Assume $\mathbi{H}$ is linear and let  
\emph{
\[
\mathcal{M}=\{\mathbi{x}_1=\cdots=\mathbi{x}_n=\mathbi{s}(t) \, | \, \mathbi{s}'(t) = \mathbi{f}(\mathbi{s}(t))\}
\]
}
be the synchronous manifold. 
Then ${\xi}$
 obeys the first variational equation 
 \emph{
\[
{\xi}' = \Big[\mathbi{I}_n\otimes D\mathbi{f}(\mathbi{s}(t)) \;-\; \alpha\big(\mathbi{L}(\mathbi{B}^{(1)})+\mathbi{L}(\mathbi{B}^{(2)})\big)\otimes H \Big]{\xi}.
\]
}
\end{proposition}

\begin{proof}
Write $\mathbi{x}_i=\mathbi{s}+{\xi}_i$, expand $\mathbi{f}$ about the solution $\mathbi{s}(t)$ and $\mathbi{H}$ at $\mathbi{0}$. Collect by pairs to obtain $\mathbi{L}(\mathbi{B}^{(1)})$ and $\mathbi{L}(\mathbi{B}^{(2)})$; the two contributions add.
\end{proof}

\subsection{Stability reduction}

\begin{theorem}[Reduction to the spectrum of the union graph]\label{thm:MSF}
Assume $\mathbi{B}^{(1)},\mathbi{B}^{(2)}$ are symmetric, and $\mathbi{H}$ is the identity. Let 
\emph{
\[
\mathbi{L}_{\mathrm{tot}} \;:=\; \mathbi{L}(\mathbi{B}^{(1)})+\mathbi{L}(\mathbi{B}^{(2)}) \;=\; \mathbi{L}\!\big(\mathbi{B}^{(1)}+\mathbi{B}^{(2)}\big).
\]}
and consider its spectrum 
$$
0=\lambda_1(\mathbi{L}_{\mathrm{tot}})<\lambda_2(\mathbi{L}_{\mathrm{tot}})\le\cdots\le\lambda_N(\mathbi{L}_{\mathrm{tot}})$$ 

Then, there is a critical coupling $\alpha_c = \alpha_c(\mathbi{f})$ such that for all 
$$
\alpha > \frac{\alpha_c}{\lambda_2(\mathbi{L}_{\mathrm{tot}})}
$$
The global synchronization is uniformly asymptotically stable
\end{theorem}

\begin{proof}
Diagonalize the symmetric $\mathbi{L}_{\mathrm{tot}}$ and project the variational system in Proposition~\ref{prop:variational} onto its eigenbasis. This yields $n$ identical $m$-dimensional systems parametrized by $\mu=\alpha\lambda_k$.
\end{proof}

\begin{remark}
The core of the above argument is that $\mathbi{L}(\mathbi{B}^{(1)}) + \mathbi{L}(\mathbi{B}^{(2)})$ has a simple spectrum; in this case, the same rationale applies. 
\end{remark}

\subsection{Fiedler monotonicity of the spectral gap}

\begin{theorem}[Fiedler monotonicity]\label{thm:Fiedler}
Let $\mathbi{W},\mathbi{W}'$ be symmetric, nonnegative $n\times n$ weight matrices on the same vertices with $\mathbi{W}'\ge \mathbi{W}$ entrywise.  Then
\[
\lambda_2(\mathbi{W}') \ge \lambda_2(\mathbi{W}).
\]
Moreover, the inequality is strict if some added edge $(i,j)$ satisfies $(W'_{ij}-W_{ij})>0$ and a Fiedler vector $\mathbi{v}$ of $\mathbi{L}(\mathbi{W})$ has $v_i\neq v_j$.
\end{theorem}

\begin{proof}
Write $\Delta:=\mathbi{W}'-\mathbi{W}\ge 0$, so $\mathbi{L}(\mathbi{W}')=\mathbi{L}(\mathbi{W})+\mathbi{L}({\Delta})$. By the Rayleigh--Ritz,
\[
\lambda_2(\mathbi{W}) = \min_{\substack{\mathbi{x}\perp \mathbf{1}\\\mathbi{x}\neq \mathbi{0}}} \frac{\mathbi{x}^\top \mathbi{L}(\mathbi{W})\mathbi{x}}{\|\mathbi{x}\|^2},\qquad
\lambda_2(\mathbi{W}') = \min_{\substack{\mathbi{x}\perp \mathbf{1}\\\mathbi{x}\neq \mathbi{0}}} \frac{\mathbi{x}^\top [\mathbi{L}(\mathbi{W})+\mathbi{L}({\Delta})]\mathbi{x}}{\|\mathbi{x}\|^2}.
\]
Since $\mathbi{x}^\top \mathbi{L}(\Delta)\mathbi{x}=\tfrac12\sum_{i,j}\Delta_{ij}(x_i-x_j)^2\ge 0$, the inequality follows. Strictness holds if some Fiedler vector $\mathbi{v}$ of $\mathbi{L}(\mathbi{W})$ satisfies $\mathbi{v}^\top \mathbi{L}(\Delta)\mathbi{v}>0$.
\end{proof}

\begin{corollary}[Connectivity inheritance]\label{cor:inherit}
Let $\mathbi{B}^{(1)},\mathbi{B}^{(2)}$ be symmetric nonnegative weighted graphs and $\mathbi{B}:=\mathbi{B}^{(1)}+\mathbi{B}^{(2)}$ their union. Then
\[
\lambda_2\!\big(\mathbi{B}\big)\;\ge\;\max\{\lambda_2(\mathbi{B}^{(1)}),\,\lambda_2(\mathbi{B}^{(2)})\}.
\]
In particular, if either $\mathbi{B}^{(1)}$ or $\mathbi{B}^{(2)}$ is connected, then $\mathbi{B}$ is connected and $\lambda_2(\mathbi{B})>0$.
\end{corollary}

\begin{proof}
Apply Theorem~\ref{thm:Fiedler} with $\mathbi{W}=\mathbi{B}^{(r)}$ and $\mathbi{W}'=\mathbi{B}$.
\end{proof}

In the case where the coupling function $\mathbi{H}$ is the identity, the hypernetwork dynamics reduce to diffusive coupling on the weighted graph with adjacency $\mathbi{B}^{(1)}+\mathbi{B}^{(2)}$. If $\mathbi{B}^{(1)}$  or $\mathbi{B}^{(2)}$  is connected, the effective Laplacian 
$\mathbi{L}(\mathbi{B}^{(1)}+\mathbi{B}^{(2)})$ has a positive Fiedler eigenvalue, and the stability of the synchronous solution is determined exactly as in the classical formalism from our previous chapters. In particular,  synchronization is guaranteed whenever one projection is connected.

\subsection{Examples on four nodes}
For each unordered triple $\{i,j,k\}$ that is a hyperedge, set $A_{ijk}=1$ for all permutations of $(i,j,k)$ and $0$ otherwise. Then $\mathbi{B}^{(1)}=\mathbi{B}^{(2)}=:\mathbi{B}$, where $B_{pq}$ equals the number of hyperedges containing the pair $\{p,q\}$.

\subsubsection*{Example U1 (two triads $\{1,2,3\}$ and $\{2,3,4\}$)}
\[
\mathbi{B}=\begin{pmatrix}
0&1&1&0\\
1&0&2&1\\
1&2&0&1\\
0&1&1&0
\end{pmatrix},\qquad
\mathbi{D}={\rm diag}(2,4,4,2),\quad \mathbi{L}=\mathbi{D}-\mathbi{B}.
\]
$\mathbi{L}$ has eigenvalues $\{0,2,4,6\}$; hence $\lambda_2(\mathbi{B})=2>0$ and the graph is connected. In \eqref{eq:model} with linear $\mathbi{H}$, the coupling equals $2$ times the Laplacian flow on $\mathbi{B}$.

\subsubsection*{Example U2 (three triads $\{1,2,3\}$, $\{1,2,4\}$, $\{2,3,4\}$)}
\[
\mathbi{B}=\begin{pmatrix}
0&2&1&1\\
2&0&2&2\\
1&2&0&1\\
1&2&1&0
\end{pmatrix},\qquad
\mathbi{D}={\rm diag}(4,6,4,4),\quad \mathbi{L}=\mathbi{D}-\mathbi{B},
\]
with eigenvalues $\{0,5,5,8\}$, so $\lambda_2(\mathbi{B})=5>0$ (stronger connectivity than U1).

\paragraph{Oriented example (both projections disconnected, union connected).}
Let $A_{ijk}=1$ on the four oriented triples
$
(1,2,3),\quad (2,1,4),\quad (3,4,1),\quad (4,3,2),
$
and $0$ otherwise. Then the projections (now generally \emph{different}) are
\[
\mathbi{B}^{(1)}=\begin{pmatrix}
0&1&0&0\\
1&0&0&0\\
0&0&0&1\\
0&0&1&0
\end{pmatrix}
\quad(\text{edges } \{1,2\},\{3,4\}),
\qquad
\mathbi{B}^{(2)}=\begin{pmatrix}
0&0&1&0\\
0&0&0&1\\
1&0&0&0\\
0&1&0&0
\end{pmatrix}
\quad(\text{edges } \{1,3\},\{2,4\}).
\]
Their sum is
\[
\mathbi{B}=\mathbi{B}^{(1)}+\mathbi{B}^{(2)}=\begin{pmatrix}
0&1&1&0\\
1&0&0&1\\
1&0&0&1\\
0&1&1&0
\end{pmatrix},
\]
which is the 4-cycle; $\mathbi{L}(\mathbi{B})$ has eigenvalues $\{0,2,2,4\}$, so $\lambda_2(\mathbi{B})=2>0$. Thus both $\mathbi{B}^{(1)}$ and $\mathbi{B}^{(2)}$ are disconnected, yet the union is connected with a positive spectral gap (Cor.~\ref{cor:inherit}).

%% file: Conclusion.tex
\motto{I have had my results for a long time: but I do not yet know how I am to arrive at them.  \\
\hspace{6cm} --- Gauss}

\chapter{Conclusions and Remarks}

We have used stability results from the theory of nonautonomous differential equations to establish  conditions for stable global synchronization in networks of diffusively coupled dissipative dynamical systems. Our conditions split the stability condition solely in terms of the isolated dynamics and network eigenvalues. 

The condition associated with the dynamics is related to the norm of the Jacobian of the vector field. This reflects the fact that to obtain stable synchronization, we need to damp all instabilities appearing in the variational equation.  The network condition is given in terms of the graph algebraic connective -- the smallest nonzero eigenvalue, which reflects how well connected the graph is.    
 
The dependence of synchronization on only two parameters is due to our hypotheses: $i)$ all isolated equations are the same, and  $ii)$ the diffusive coupling between them is mutual. These assumptions allow for rigorous results. 

There are other approaches to tackling the stability of the global synchronization. Successful approaches are the construction of a Lyapunov function of the synchronization manifold, see  for example Refs. \cite{Henk,Hasler,Li}, which takes a control view; and the  theory of invariant manifolds \cite{Josic1,Josic2} taking a dynamical system view. 
Our results have a deeper connection with the previous approach introduced by Pecora and Carrol \cite{Pecora1}. They used the theory of Lyapunov exponents, which allows the tackling of general coupling functions. The main drawback is that of obtaining results for the persistence of the global synchronization. This requires establishing results on the continuity of the Lyapunov exponent, which is rather subtle  \cite{EDO1}. \footnote{Small perturbations 
can destabilize a system with negative Lyapunov exponents. To guarantee the persistence under 
perturbations, Lyapunov regularity is required, see Ref. \cite{EDO1}. } 

The approach introduced in these notes follows the steps of the Pecora and Carrol analysis, that is, the local stability analysis of the synchronization manifold, but uses various concepts in stability theory, to establish the persistence results for the global synchronization.
We also left out the treatment when the graph is directed. With some extra effort, directed graphs can be treated with similar techniques. This case is interesting as some improvements in the graph topology can hinder synchronization \cite{Improving,PRE_Adding}

%% file: appendix_EDO.tex
%
%
%

\appendix
\motto{If only I had the theorems! Then I should find the proofs easily enough. \\
\hspace{5cm} -- Bernard Riemann}

\chapter{Linear Algebra}\label{LA}

For this exposition we consider the field $F$ where $F =\mathbb{R}$ the field
of real numbers or $F = \mathbb{C}$ the field of complex numbers. We shall closely 
follow the exposition of Ref. \cite{Lancaster}. Consider the set Mat($F,n$) of all square matrices acting on $F^{n}$.
We start with the following

\begin{Definicao} Let $\mathbi{A} \in $ \mbox{  Mat( $F,n $ )}.  The set 
$$
\sigma(\mathbi{A}) : = \{ \lambda \in \mathbb{C} : \det(\mathbi{A} - \lambda \mathbi{I}) = 0\}
$$
is called the spectrum of $\mathbi{A}$. 
\end{Definicao}

The spectrum of $\mathbi{A}$ is constituted of all its eigenvalues. Note by the fundamental 
theorem of algebra the spectrum has at most $n$ distinct points. 


Often, we want to obtain bounds on the localization of eigenvalues on the 
complex plane. A handy result is provided by the result
 
\begin{Teorema}[Gershgorin]
Let $ \mathbi{A} \in Mat (\mathbb{C},n)$, denote $\mathbi{A}  = (A_{ij})_{i,j=1}^n$. Let $D(a, \delta)$ 
denote the ball of radius $\delta$ centered at $a$. 
For each $1 \le i \le n$ let 
$$
R_i = \sum_{j=1 \atop{j\not=i}}^n |A_{ij}|, 
$$
then every eigenvalue of $\mathbi{A}$ lies within at least one of the balls $D(A_{ii}, R_i)$.
\label{LA:ThmG}
\end{Teorema}

For a proof see Ref. \cite{Lancaster} Sec. 10.6.

If $ \mathbi{A} \in Mat (\mathbb{C},n)$ then we denote its conjugate transpose by $ \mathbi{A}^*$. 
In case $\mathbi{A}$ is a real valued matrix $\mathbi{ A}^*$ denotes the transpose. A matrix 
is called hermitian if $\mathbi{A} =  \mathbi{A}^*$. 
The following definition is also fundamental 
 
\begin{Definicao}
Let $\mathbi{A} \in Mat (\mathbb{C},n)$ be a  hermitian matrix.  
It is called positive-semidefinite (or sometimes nonnegative-definite) if
$$
\mathbi{x}^* \mathbi{A} \mathbi{x} \ge 0
$$
for any $\mathbi{x} \in \mathbb{C}^n$
\end{Definicao}
It follows that a matrix is nonnegative if all its eigenvalues are non negative. 

\section{Matrix space as a normed  vector Space}

Consider the vector space $F^{n}$ over the field $F$. A norm $\| \cdot \|$ on 
$F^n$ is a function $\| \cdot \| : F^n \rightarrow \mathbb{R}$ satisfying
\begin{enumerate}
\item positive definiteness : $\| \mathbi{x} \| \ge 0$ for all $\mathbi{x} \in F^n$ and equality holds iff $\mathbi{x} = \mathbi{0} $
\item Absolute  definiteness : $\| \gamma \mathbi{x} \| = |\gamma| \| \mathbi{x} \|$ for 
all $\mathbi{x} \in F^n$ and $\gamma \in F$ 
\item Triangle inequality : $\| \mathbi{x} + \mathbi{y} \| \le \| \mathbi{x} \| + \| \mathbi{y}\|$ for all $\mathbi{x},\mathbi{y} \in F^n$
\end{enumerate}

We call the pair ($F^n, \| \cdot \|$) is called normed vector space. 
This normed vector space is also a metric space under the metric $d: F^n \times F^n \rightarrow \mathbb{R}$
where $d(\mathbi{x} , \mathbi{y} ) = \| \mathbi{x} - \mathbi{y} \|$. We say that  $d$ is the metric induced by the norm.
In this metric, the norm defines a continuous map from $F^n$ to $\mathbb{R}$, and the norm 
is a convex function of its argument.  Normed vector spaces are central to the study of linear algebra.

Once we introduce of norm on the vector space $F^n$, we can also view 
the Mat($F,n$) as a normed spaces. This can be done by the induced matrix norm 
which is a natural extension of the notion of a vector norm  to matrices.
Given a vector norm $\| \cdot \|$ on $F^n$, we define  the corresponding induced 
norm or operator norm on the space Mat($F,n$) as:
$$
\| \mathbi{A} \| = \sup \{ \| \mathbi{A x} \| : \mathbi{x}\in F^n \mbox{ and  }  \| \mathbi{x} \| =1 \}
$$

It follows from the theory of functions on compact spaces that $\| \mathbi{A} \| $ always exists and 
it is called induced norm. Indeed, the induced norm defines defines a norm on Mat($F,n$) 
satisfying the properties 1-3  and an additional property
$$
\| \mathbi{A B} \| \le \| \mathbi{A}  \| \| \mathbi{B} \| \mbox{  for all  } \mathbi{A},\mathbi{B} \in \mbox{  Mat($F,n$)}
$$
called sub-multiplicativity. A  sub-multiplicative norm 
on   Mat($F,n$) is called matrix norm or operator norm.  Note that even though we 
use the same notation $\| \mathbi{A} \| $ for the norm of $A$, this should not be confused 
with the vector norm. 

\begin{ex}\label{norInf}
Consider the norm of the maximum $\| \cdot \|_{\infty}$ on $\mathbb{R}^n$. Given 
$\mathbb{R}^n \ni \mathbi{x} = (x_1, \cdots, x_n)$,  the norm is defined as
$\| \mathbi{x} \| = \max_i |x_i |$. Given a matrix $\mathbi{A} = (A_{ij})_{i,j=1}^n$
then
$$
\| \mathbi{A} \|_{\infty} = \max_{1 \le i \le n} \sum_{j=1}^n |A_{ij}|
$$
\end{ex}

\begin{ex}
Consider the Euclidean norm $\| \cdot \|_{2}$ on $\mathbb{R}^n$. Using the notation 
of the last example, we have
$$
\| \mathbi{A} \|_{2} =  \sqrt{\rho(\mathbi{A}^* \mathbi{A})}
$$
where $\rho_{\max} (\mathbi{A}^* \mathbi{A})$ is spectral radius $\mathbi{A}^* \mathbi{A}$.
\end{ex}

Recall that two norms $\| \|^{\prime}$ and $\| \|^{\prime \prime}$ are said to be equivalent if 
$$
a \| \mathbi{A} \|^{\prime} \le \| \mathbi{A}\|^{\prime \prime} \le b \| \mathbi{A} \|^{\prime}
$$
for some positive numbers $a,b$ and for all matrices $\mathbi{A}$.
It follows that in finite-dimensional normed vector spaces any two norms are equivalents. 

\section{Representation Theory}

We review some fundamental results on matrix representations.
A square matrix $\mathbi{A}$ is diagonalizable if and only if there exists a basis of 
$F^n$ consisting of eigenvectors of $\mathbi{A}$. In other words, if the $F^n$ is 
spanned by the eigenvectors of $\mathbi{A}$. If such a basis can be found, then 
$\mathbi{P}^{-1} \mathbi{A} \mathbi{P}$ is a diagonal matrix, where $\mathbi{P}$ is the eigenvector matrix, 
each column of $\mathbi{P}$ consists of an eigenvector. The diagonal entries of this 
matrix are the eigenvalues of $\mathbi{A}$. One of the main goals in matrix analysis is to 
classify the diagonalizable matrices. 

In general diagonalization will depend on the properties of $F$ such as whether $F$ is a
algebraically closed field. If $F = \mathbb{C}$ then almost every matrix is diagonalizable. 
In other words, the set $B \subset$ Mat ($\mathbb{C},n$) of non diagonalizable matrices 
over $\mathbb{C}$ has Lebesgue measure zero. Moreover, the set diagonalizable matrices 
form a dense subset. Any non diagonalizable matrix, say $\mathbi{Q} \in B$ can be 
approximated by a diagonalizable matrix. Precisely,  given $\varepsilon > 0$ there is a sequence
$\{ \mathbi{A}_i \}$ of diagonalizable matrices such that $\| \mathbi{Q} - \mathbi{A}_i \| < \varepsilon $ 
for any $i > n_0$.

Let us denote by $^*$ the conjugate transpose if $F = \mathbb{C}$ (clearly only 
transpose if $F = \mathbb{R}$). We first focus on symmetric matrices $\mathbi{A} = \mathbi{A}^{*}$ 
and $F = \mathbb{R}$. It turns out that it is always possible to diagonalize such matrices. 

\begin{Definicao} A real square matrix $\mathbi{A}$ is orthogonally diagonalizable if 
there exists an orthogonal matrix $\mathbi{P}$ such that $\mathbi{P}^* \mathbi{A} \mathbi{P} = \mathbi{D}$ 
is a diagonal matrix.
\end{Definicao}

Diagonalization of symmetric matrices is guaranteed by the following

\begin{Teorema} \label{LA:ThmDiag}
Let $\mathbi{   A}$ be a real symmetric matrix. Then there exists an orthogonal matrix $\mathbi{   P}$ such that :
\begin{enumerate}
\item $\mathbi{P}^* \mathbi{A} \mathbi{P}  = \mathbi{D}$ is  a diagonal matrix.
\item $\mathbi{D} =  \mbox{diag }(\lambda_1, \cdots , \lambda_n)$, where $\lambda_i$ are the eigenvalues of $\mathbi{A}$. 
\item The column vectors of $\mathbi{P}$ are the eigenvectors of the eigenvalues of $\mathbi{A}$.
\end{enumerate}
\end{Teorema}

For a proof see Ref. \cite{Golub} Sec. 8.1.

\section{Kronecker Product}

We need several properties of the Kronecker Product to address the 
stability of the synchronized motion in networks. 
\begin{Definicao}
Let $\mathbi{A} \in $ Mat($F, m \times n$) and $\mathbi{B} \in $ Mat($F, r \times s$). The 
Kronecker Product of the matrices $A$ and $B$ and defined as the matrix
$$
\mathbi{A} \otimes \mathbi{B} = 
\left(
\begin{array}{ccc}
A_{11} \mathbi{B} & \cdots &  A_{1n} \mathbi{B} \\
\vdots & \ddots & \vdots \\ 
A_{m1} \mathbi{B}  & \cdots & A_{mn} \mathbi{B} 
\end{array}
\right)
$$
\end{Definicao}

The Kronecker product is sometimes called tensor product. Consider now the following 
examples on the 
\begin{ex}
Consider the matrices
$$\mathbi{A} = \left(
\begin{array}{cc}
a & b \\
c & d
\end{array}
\right) 
\mbox{  and }
\mathbi{B} = \left(
\begin{array}{cc}
1 & 0 \\
2 & 3
\end{array}
\right)
$$
Then 
$$\mathbi{A} \otimes \mathbi{B} =  \left(
\begin{array}{cc}
a \mathbi{B}& b \mathbi{B}\\
c \mathbi{B} & d \mathbi{B} 
\end{array}
\right) = 
\left(
\begin{array}{cccc}
a & 0& b & 0 \\
2 a & 3 a& 2 b & 3 b \\
c & 0 a& d & 0  \\
2 c & 3c & 2d  & 3 d \\
\end{array}
\right) 
$$
Now consider the vectors
$$\mathbi{v} = \left(
\begin{array}{c}
1 \\
1
\end{array}
\right) 
\mbox{  and }
\mathbi{u}(t) = \left(
\begin{array}{cc}
x(t) \\
y(t)
\end{array}
\right)
$$
Then 
$$\mathbi{v} \otimes \mathbi{ u}(t) =  \left(
\begin{array}{c}
x(t) \\
y(t) \\
x(t) \\
y(t) 
\end{array}
\right)
$$
\end{ex}

We review the basic results we need. 

\begin{Teorema} \label{Thm:KroPod}
Let $\mathbi{A} \in $ Mat($F, m \times n$) and $\mathbi{B} \in $ Mat($F, r \times s$)
$\mathbi{C} \in $ Mat($F, n \times p$) and $\mathbi{D} \in $ Mat($F, s \times t$). 
Then 
$$
( \mathbi{A} \otimes \mathbi{B})(\mathbi{C} \otimes \mathbi{D})=\mathbi{A}\mathbi{C} \otimes \mathbi{B}\mathbi{D}.
$$
\end{Teorema}

The proof can be found in Ref. \cite{Lancaster} pg. 408, see Proposition 2.  Note that 
$( \mathbi{A} \otimes \mathbi{B})(\mathbi{C} \otimes \mathbi{D}) \in $ Mat ($F, mr \times pt$).
A direct computation leads to the following result

\begin{Teorema}\label{Thm:KroT}
Let $\mathbi{A} \in $ Mat($F, m \times n$) and $\mathbi{B} \in $ Mat($F, r \times s$), then 
$$
(\mathbi{A} \otimes \mathbi{B})^* = \mathbi{A}^* \otimes \mathbi{B}^*
$$
\end{Teorema}

By applying Theorem \ref{Thm:KroPod} we conclude that following

\begin{Teorema} \label{Thm:Inv} If \mathbi{A}  and \mathbi{B} are nonsingular, then
$$
(\mathbi{A} \otimes \mathbi{B})^{-1} = \mathbi{A}^{-1} \otimes \mathbi{B}^{-1}.
$$
\end{Teorema}

We following Theorem also plays a important role in the exposition

\begin{Teorema} Let $\{ \lambda_i \}_{i=1}^r$ be the eigenvalues of $\mathbi{A} \in $ Mat($F, n$) 
and $\{ \mu_i \}_{i=1}^s$ be the eigenvalues of $\mathbi{  B} \in $ Mat($F, n$). 
Then $ \mathbi{A} \otimes \mathbi{B}$ has $rs$ eigenvalues 
$$
\lambda_1 \mu_1,...,\lambda_1 \mu_s, \lambda_2 \mu_1, \cdots ,\lambda_2 \mu_s,...,\lambda_r \mu_s.
$$
\end{Teorema}

The proof can be found in Ref. \cite{Lancaster} pg. 412.  A direct consequence of this result is the following

\begin{Teorema}\label{Thm:KroP}
Let $\mathbi{A}$ and $\mathbi{B}$ be positive semi-definite matrices. 
Then $\mathbi{A}  \otimes \mathbi{B} $ is also positive semi-definite.
\end{Teorema}

Our last result concerns the norms of the Kronecker products 

\begin{Teorema}\label{normK} Let $\|  \cdot \|_p$ be $p$-norm. Consider $\mathbi{v} \in \mathbb{R}^s$, 
and  $\mathbi{x} \in \mathbb{R}^t$, for $t,s \in \mathbb{N}$. Then 
$$
\| \mathbi{v} \otimes \mathbi{x} \|_p = \| \mathbi{v} \|_p \| \mathbi{x} \|_p 
$$
\end{Teorema}

\chapter{Ordinary Differential Equations} \label{DE}

Let $D$ be an open connected subset 
of $\mathbb{R}^m$, $m \ge  1$, and let $\mathbi{G }:  D \rightarrow \mathbb{R}^m$ be an
autonomous vector field. 
Consider the problem of finding solutions for the vector differential equation 
\begin{equation}
\dot{\mathbi{x}} = \mathbi{G}(\mathbi{x})
\label{G}
\end{equation}
with the initial condition $\mathbi{x}(0) = \mathbi{x}_0$. A positive answer to this problem is given 
by the following

\begin{Teorema} [Picard-Lindel\"of] Assume that the vector field $\mathbi{G}$ Lipschitz continuous 
in a neighborhood of $\mathbi{x}_0$. Precisely, assume that given $\mathbi{x}_0 \in  U \subset D$ there 
is a constant $K_U$ such that 
$$
\| \mathbi{G}(\mathbi{x}) - \mathbi{G}(\mathbi{y}) \| \le K_U \|  \mathbi{x} - \mathbi{y} \|
$$
for all $\mathbi{   x},\mathbi{   y} \in U$. 
Then there exists a unique local solution $\mathbi{x}(t)$ for Eq. (\ref{G}) satisfying 
$\mathbi{x}(0) = \mathbi{x}_0$.
\label{ThmPL}
\end{Teorema}

Note that the solution is local, in the sense that there is small $\kappa>0$ such that the 
function $\mathbi{x} : [-\kappa,\kappa] \rightarrow D$ is a solution of the problem with 
$\mathbi{x}(0) = \mathbi{x}_0$. The question is: How long does such solution exist for? 
We are interested in the long term behavior of the solutions, so we wish to know
under what conditions the solutions exists  forward in time. A positive answer is given by 
extension theorems:

\begin{Teorema} [Extension]
Let $\mathcal{C}$ be a compact subset of the open set $D$.
Consider Eq. (\ref{G}) and let $\mathbi{G}$ be differentiable.   Let $\mathbi{x}_0 \in \mathcal{C}$ 
and suppose that every solution $\mathbi{x} : [0, \tau ] \rightarrow D$ with $\mathbi{x}(0) = \mathbi{x}l_0$ lies 
entirely in $C$. Then this solution is defined for all (forward) time  $t \ge0$. 
\label{ThmExt}
\end{Teorema}

The proofs of the above theorems can be founds in Refs. \cite{EDOHartman,Smale}.

\section{Linear Differential Equations}

The evolution operator also determines the behavior of the non homogeneous equation

\begin{Teorema} Let $\mathbi{A} : \mathbb{R} \rightarrow $ Mat($\mathbb{R},n$) and 
$\mathbi{g} : \mathbb{R} \rightarrow \mathbb{R}^n$  be 
continuous function. 
Consider the perturbed equation
$$
\mathbi{y} = \mathbi{A} \mathbi{y} + \mathbi{g}(t)
$$
The solution of the perturbed equation corresponding to the initial condition 
$\mathbi{x}(t_0) = \mathbi{x}_0$ is given by
$$
\mathbi{y}(t) = \mathbi{T}(t, t_0)\mathbi{x}_0 + \int_{t_0}^t \mathbi{T}(t,s) \mathbi{g}(s) ds 
$$
where $\mathbi{T}(t, t_0)$ is the evolution operator of the corresponding homogeneous 
system.
\label{ThmVP}
\end{Teorema}

The following inequality is central to obtain various estimates

\begin{Lema}[Gronwall]  Consider $U \subset \mathbb{R}_+$ and 
let $u : U \rightarrow \mathbb{R}$ be continuous and nonnegative function.
Suppose there exist $C \ge 0$ and and $K \ge 0$ such that
\begin{equation}
u(t) \le C + \int_0^t K u(s)ds
\label{ules}
\end{equation}
for all $t \in U$, then
$$
u(t) \le C e^{Kt}.
$$
\label{ThmGI}
\end{Lema}

The proof of these results can be found in Ref. \cite{EDOHartman}.

%% file: LectureNotes.bbl
\begin{thebibliography}{99}


\bibitem{Newman} M. E. J. Newman, {\it Networks: An Introduction}, Oxford University Press (2010).

\bibitem{Albert1} R.~Albert, H.~Jeong, A.-L.~Barab\'asi A.-L. , Nature { 406}, 378 (2000).


\bibitem{Albert2} R.~Albert, A.-L.~Barab\'asi, Rev. Mod. Phys. { 74}, 47 (2002).


\bibitem{Net3} D.L.~Turcotte, {\it Fractals and Chaos in Geology and Geophysics}, 2nd edn, Cambridge UP, 1997.

\bibitem{Net4} R.M.~May, {\it Stability and Complexity in Model Ecosystems}, Princeton UP, 1973. 

\bibitem{Net5} S.A.~Levin, B.T.~Grenfell, A.~Hastings, A.S.~Perelson, 
Science 275, 334–343 (1997). 

\bibitem{Bullmore} E. Bullmore, O. Sporns,  Nature Neurosc. { 10}, 186 (2009).


\bibitem{neutrinos} J.~Pantaleone, Phys. Rev. D 58, 3002 (1998).

\bibitem{StrogatzJunction}  K.~Wiesenfeld, P.~ Colet, and S.  Strogatz,  Phys. Rev. E 57, 1563 (1998).



\bibitem{Net1} A.T.~Winfree,  {\it The Geometry of Biological Time}, Springer, 1980.

\bibitem{Net2} Y.~Kuramoto, {\it Chemical Oscillations, Waves, and Turbulence}, Springer, 1984. 

\bibitem{Fries} P.~Fries, Trends Cogn.Sci. 9, 474 (2005) .

\bibitem{HubSync} T. Pereira, Phys. Rev. E {\bf 82}, 036201 (2010).

\bibitem{RMP}
T Stankovski, T Pereira, PVE McClintock, A Stefanovska, 
Reviews of Modern Physics 89, 045001 (2017)


\bibitem{Strogatz} S. Strogatz,{\it Sync: The Emerging Science of Spontaneous Order}, 
Hyperion, New York, (2003).   

\bibitem{Kurths} A. Arenas, A. Diaz-Guilera, J. Kurths, Y. Moreno, C. Zhou
{\it Synchronization in complex networks}, Physics Rep. {\it 469}, 93 (2008).

\bibitem{Wu} C. W. Wu, {\it Synchronization in complex networks of nonlinear dynamical systems}, 
World Scientific Publishing Co. Pte. Ltd., Singapore (2007).

\bibitem{Mech} H. Nijmeijer and A. Rodr\'{i}guez-Angeles, {\it Synchronization of mechanical systems},
World Scientific Publishing Co. Pte. Ltd., Singapore (2003).


\bibitem{Tiago} T Pereira, MS Baptista, J Kurths, Phys. Rev. E 75, 026216 (2007).	

\bibitem{TiagoPhD_2006} T Pereira, MS Baptista, J Kurths,  Physica D 216, 260 (2006).

\bibitem{DenizSync_Review} D Eroglu, JSW Lamb, T Pereira, 
Contemporary Physics 58, 207-243 (2017).


\bibitem{WolfSinger} C.M.~Gray, W. Singer, Proc. Nat. Acad. Sci. USA { 86}, 1698 (1989).

\bibitem{Attention}  G.G.~Gregoriou, S.J.~Gotts, H.~Zhou, R.~Desimone Science { 324}, 1207 (2009).


\bibitem{Singer} W.~Singer,  Neuron { 24}, 49 (1999).

\bibitem{Ep} John Milton and Peter Jung (Ed), {\it 
Epilepsy as a Dynamic Disease}, Springer, 2010.

\bibitem{PRX} D Eroglu, M Tanzi, S van Strien, T Pereira, 
Physical Review X 10, 021047 (2020)


\bibitem{parkinson} P.~Tass, M.G.~Rosenblum, J.Weule, {\it et al.}, Phys.Rev.Lett. { 81}, 3291 (1998).

\bibitem{Extinction} D.J.D.~Earn, S.A.~Levin, P.~Rohani, Science 290, 1360 (2000).

\bibitem{Spread} B.T.~Grenfell {\it et al.}, Nature 414, 716 (2001).



\bibitem{Lai-Sang1} S Ruschel, T Pereira, S Yanchuk, LS Young, 
Journal of Mathematical Biology 79, 249-279 (2019)

\bibitem{Lai-Sang2} LS Young, S Ruschel, S Yanchuk, T Pereira, 
Scientific Reports 9, 3505 (2019)

\bibitem{Pecora1}  L.M.  Pecora and  T.L. Carrol, Phys. Rev. Lett. {\bf 80}, 2109 (1998); 

\bibitem{Explosive} V Vlasov, Y Zou, T Pereira
Physical Review E 92 (1), 012904 (2015) 


\bibitem{Comb} F. R. K. Chung and L. Lu, {\it Complex Graphs and Networks}, American Mathematical Society (2006).


\bibitem{Barrat} A. Barrat, M. Barthelemi, A. Vespegnani, {\it Dynamical Processes 
on Complex Networks}, Cambridge University Press (2008).

\bibitem{JEMS} T Pereira, S van Strien, M Tanzi
Journal of the European  Mathematical Society 22, 2183–2252 (2020) 

\bibitem{StructuralGen} C Poignard, T Pereira, JP Pade, 
SIAM Journal on Applied Mathematics 78, 372-394 (2018)


\bibitem{JNLS} C Poignard, JP Pade, T Pereira, 
Journal of Nonlinear Science 29 (5), 1919-1942 (2019)




\bibitem{Pecora2} M. Barahona and L.M. Pecora, Phys. Rev. Lett. {\bf 89}, 054101 (2002).

\bibitem{Motter} A.E. Motter, C. Zhou, and J. Kurths, Phys. Rev. E {\bf 71}, 
016116 (2005).

\bibitem{NonlinChimera} J Eldering, JSW Lamb, T Pereira, ER dos Santos, Nonlinearity 34 (8), 5344 (2021) 


\bibitem{ZhengChaos} RM Corder, Z Bian, T Pereira, A Montalbán, 
Chaos, 33,  091103 (2022)

\bibitem{ZhengCMP} Z Bian, JSW Lamb, T Pereira, 
Communications in Mathematical Physics 406, 170 (2025). 


\bibitem{PRL_T} T. Pereira, D. Eroglu, G. B. Bagci, U. Tirnakli, H. J. Jensen,  Phys. Rev. Lett. 110, 234103 (2013).

\bibitem{Jaap} T. Pereira, J. Eldering, M. Rasmussen, A. Veneziani, {\it Towards a general theory for coupling functions allowing persistent synchronization}, Nonlinearity 27 (3), 501 (2014).


\bibitem{LyapB} L. Barreira and Y.B. Pesin , {\it Lyapunov Exponents and Smooth Ergodic Theory}, American Mathematical Society (2002).  


\bibitem{Graph} B. Bollobas, {\it Modern Graph Theory}, Springer (1998).

\bibitem{Mohar} Bojan Mohar, GRAPHS AND COMBINATORICS
Volume 7, Number 1, 53-64, DOI: 10.1007/BF01789463

\bibitem{Fiedler} M. Fiedler, Algebraic connectivity of graphs, Czech. Math. J. 23 (98) (1973) 298–305.

\bibitem{Bojan} B. Mohar, Graph Theory, Comb. Appl.  {\bf 2}, 871 (1991). 

\bibitem{Bojan2} B. Mohar,  {\it Graph Symmetry: Algebraic Methods and Applications}, NATO 
ASI Series C vol. {\bf 497 } (1997),  pgs. {227--275}.


\bibitem{Liapunov} A. Bacciotti and L. Rosier, {\it Liapunov Functions and Stability in Control Theory}, 
Springer-Verlag Berlin Heidelberg (2005).

\bibitem{nonlin} A.~Katok, B.~Hasselblatt, {\it 
Introduction to the Modern Theory of Dynamical Systems}, Cambridge UP, 1996. 


\bibitem{Viana} M. Viana, {\it What's new on Lorenz strange attractors?} Math. Intelligencer, {\bf  22}, 6 (2000). 

\bibitem{Sparow} C. Sparow, {The Lorenz Equations: Bifurcations, Chaos, and Strange Attractors}, 
Springer (1982) 

\bibitem{DiffusionDriven} A. Pogromsky, T. Glad, and H.  Nijmeijer, Int. J. Bif. Chaos  9,  629 (1999).

\bibitem{CalculusRn} R. Courant, D. Hilbert. {\it Methods of
Mathematical Physics}. Vol 1. Interscience Publishers,
Inc. New York, 1953.

\bibitem{Coppel}  W.A. Coppel, {\it Dichotomies in Stability Theory}, Springer-Verlag Berlin Heidelberg New York (1978).

\bibitem{LinearSys}  P. J. Antsaklis and A. N. Michel, {\it Linear Systems}, Mcgraw-Hill College (1997)

\bibitem{rasmussen} M. Rasmussen,  {\it Attractivity and Bifurcation for Nonautonomous Dynamical Systems}, Lecture Notes in Mathematics 1907, Springer (2007).

\bibitem{Henk} A. Pogromsky and H. Nijmeijer,  IEEE Trans. Circ. Sys. - I 48,  152 (2001).

\bibitem{Hasler} V. Belykh, I. Belykh and M. Hasler, Physica D 195,  159 (2004).

\bibitem{Li} C-H. Li and S-Y Yang,  J. London Math. Soc. (2011); {\sf doi:10.1112/jlms/jdq096}.

\bibitem{Josic1} K. Josic,  Phys. Rev. Lett. 80, 3053 (1998).

\bibitem{Josic2} K. Josic, Nonlinearity 13,  1321 (2000). 

\bibitem{EDO1} L.  Barreira  and C. Valls, {\it Stability of Nonautonomous 
Differential Equations}, Springer-Verlag Berlin Heidelberg (2008).  

\bibitem{Lancaster}  P. Lancaster and M. Tismenetsky ,  {\it The Theory of Matrices},
Academic Press; 2 edition (1985).

\bibitem{Golub} G. H. Golub, C. F. Van Loan, {\it  Matrix Computations}, The Johns Hopkins University 
Press; 3rd edition (1996).


\bibitem{EDOHartman} P. Hartman,  {\it Ordinary Differential Equations},  John Wiley \& Sons, Inc,  NY (1964).

\bibitem{Smale} M. W. Kirsch, S. Smale, e R.L. Devaney, {\it Differential Equations, Dynamical Systems and An Introduction to Chaos}, Academic Press, San Diego (2004).

\bibitem{EDOHartman} P. Hartman,  {\it Ordinary Differential Equations},  John Wiley \& Sons, Inc,  NY (1964).







\bibitem{EddieNatCom} E Nijholt, JL Ocampo-Espindola, D Eroglu, IZ Kiss, T Pereira, Nature communications 13 (1), 4849 (2022)

\bibitem{EddieHyper} S Von Der Gracht, E Nijholt, B Rink, 
SIAM Journal on Applied Mathematics 83 (6), 2329-2353 (2023)

\bibitem{Bick} C Bick, E Gross, HA Harrington, MT Schaub
SIAM review 65 (3), 686-731

\bibitem{CoheRalf} R Tönjes, CE Fiore, T Pereira,
Nature Communications 12 (1), 72 (2021)

\bibitem{Improving} JP Pade, T Pereira, 
Scientific Reports 5 (1), 9968 (2015)



\bibitem{PRE_Adding} JD Hart, JP Pade, T Pereira, TE Murphy, R Roy, 
Physical Review E 92 (2), 022804 (2015).

\end{thebibliography}
